\documentclass[a4paper,fleqn,usenatbib,useAMS]{mnras}

\usepackage[T1]{fontenc}
\usepackage{ae,aecompl}

\usepackage{graphicx}	
\usepackage{amsmath}	
\usepackage{amssymb}	

\usepackage{times}
\usepackage{epstopdf}
\usepackage{txfonts}
\usepackage{natbib}
\usepackage{array}
\usepackage{verbatim}
\usepackage{subfig}
\usepackage{stackengine}
\title[UV-continuum in Lyman-$\alpha$ emitters]{
Resolving on 100~pc-scales the UV-continuum in Lyman-$\alpha$ emitters between redshift 2 to 3 with gravitational lensing}

\author[E. Ritondale et al.]{
E. Ritondale,$^{1}$\thanks{E-mail: elisa@mpa-graching.mpg.de} 
M.~W. Auger,$^{2}$ 
S. Vegetti$^{1}$ and
J.~P. McKean$^{3,4}$
\\
$^1$Max Planck Institute for Astrophysics, Karl-Schwarzschild-Strasse 1, D-85748 Garching, Germany\\
$^2$Institute of Astronomy, University of Cambridge, Madingley Rd, Cambridge CB3 0HA, United Kingdom\\
$^3$ASTRON, Netherlands Institute for Radio Astronomy, Oude Hoogeveensedijk 4, 7991 PD Dwingeloo, the Netherlands\\
$^4$Kapteyn Astronomical Institute, University of Groningen, Postbus 800, NL$-$9700 AV Groningen, the Netherlands
}

\date{Accepted XXX. Received YYY; in original form ZZZ}

\pubyear{2018}

\begin{document}
\label{firstpage}
\pagerange{\pageref{firstpage}--\pageref{lastpage}}
\maketitle

\begin{abstract}
We present a study of seventeen LAEs at redshift 2$<z<$3 gravitationally lensed by massive early-type galaxies (ETGs) at a mean redshift of approximately 0.5. Using a fully Bayesian grid-based technique, we model the gravitational  lens mass distributions with elliptical power-law profiles and reconstruct the UV-continuum surface brightness distributions of the background sources using pixellated source models. We find that the deflectors are close to, but not consistent with isothermal models in almost all cases, at the $2\sigma$-level. We take advantage of the lensing magnification (typically $\mu\simeq$ 20) to characterise the physical and morphological properties of these LAE galaxies. From reconstructing the ultra-violet continuum emission, we find that the star-formation rates range from 0.3 to 8.5~M$_{\odot}$~yr$^{-1}$ and that the galaxies are typically composed of several compact and diffuse components, separated by 0.4 to 4 kpc. Moreover, they have peak star-formation rate intensities that range from 2.1 to 54.1~M$_{\odot}$~yr$^{-1}$~kpc$^{-2}$. These galaxies tend to be extended with major axis ranging from 0.2 to 1.8 kpc (median 561~pc), and with a median ellipticity of 0.49. This morphology is consistent with disk-like structures of star-formation for more than half of the sample. However, for at least two sources, we also find off-axis components that may be associated with mergers. Resolved kinematical information will be needed to confirm the disk-like nature and possible merger scenario for the LAEs in the sample.

\end{abstract}

\section{Introduction}
Lyman-$\alpha$ emitting (LAE) galaxies represent a population of star-forming systems with very large Lyman-$\alpha$ equivalent widths and have some of the highest known specific star-formation rates (sSFR) in the Universe \citep*[e.g.][]{Hu98}. Typical LAE galaxies have strong star-formation, high-ionisation, and are typically low metallicity and low mass galaxies \citep[e.g.][]{Erb06,Gawiser06}. These properties, combined with a (mostly) low-dust content, allow for the escape of a significant fraction of Lyman-$\alpha$ photons \citep[e.g.][]{Gronwall07,Hayes10,Marques17}. Therefore, these low-mass galaxies may be an essential population as they are thought to be predominantly responsible for the re-ionisation of the Universe \citep[e.g.][]{Malhotra04,Nakajima16}. 

At redshift $2 < z < 3$, well-studied LAEs from wide-field surveys are typically at the bright end of this parameter space, being  $L^\text{*}$ galaxies with $M^\text{*} \sim$ 10$^9$ M$_\odot$ and typical SFRs of about a few to 100 M$_\odot$~year$^{-1}$ \citep[e.g.][]{Gawiser06,Erb14}. Investigations of lower-SFR objects have generally been limited to quantifying the properties of strong optical lines \citep[e.g.][]{Trainor15,Trainor16}, although deep narrow-band imaging has also uncovered a large population of both low-mass and low SFR LAEs \citep[e.g.][]{Shimakawa17}. For example, it has been recently shown that low-SFR LAEs, with similar characteristics to the local-Universe \emph{green peas} (M$^\text{*}$ as low as $10^7$~M$_\odot$ and SFR of about 1 to 100~M$_\odot$~year$^{-1}$), have strong optical emission line (H$\alpha$ and [O\,{\sc iii}]) properties that are consistent with optically-selected star-forming galaxies of the same stellar masses at $z \sim 2$ \citep{Hagen16}. However, it is not possible to directly determine the gas metallicity, density, and kinematics of these galaxies without substantial investments in telescope time. 
Finally, high-resolution imaging studies find that LAEs are typically compact, with no evidence for strong evolution in size with redshift \citep[e.g.][]{Venemans05,Malhotra12,Paulino-Afonso17}. Their Lyman-$\alpha$ halo, which is typically more extended than the UV-continuum by a factor of 10 in average, has also been found not to evolve with cosmic time \citep{Leclercq17}. However, such studies are currently limited by the angular resolution of the observations.

In principle, strong gravitational lensing can be used to overcome these limitations. In practice, however, most of the strongly lensed galaxies at $z \sim 2$ with moderate star formation are not LAEs \citep[e.g.][]{Hainline09,Rhoads14,Stark13}, and at present, the properties of only a few lensed LAEs could be investigated in detail \citep{Chris12, Vanzella16,Patricio16}. Recently, new {\it Hubble Space Telescope} ({\it HST}) V-band observations of LAE galaxies selected from the Baryon Oscillation Spectroscopic Survey (BOSS) have revealed a sample of strongly lensed systems at $\langle z\rangle \sim$ 2.5. Thanks to the lensing magnification we can probe the detailed structure of these galaxies at scales around 100~pc \citep{Shu16a}.

In this paper, we use strong gravitational lensing to go beyond the current limits in angular resolution and investigate the size and structure of LAE galaxies at redshift 2$<z<$3 on 100--500 pc-scales. Our study focuses on the first statistically significant sample of strong gravitational lenses with high-redshift LAEs as their background sources that were selected from the BOSS Emission Line Lens Survey (BELLS) by \citet{Shu16a}. To summarise, $1.4\times10^6$ galaxy spectra from the BOSS survey of the Sloan Digital Sky Survey-III were inspected to search for Lyman-$\alpha$ emission lines at a higher redshift than the dominant early-type galaxy in the spectrum. From this search, \citet{Shu16a} selected twenty-one highest quality targets with source redshifts between $z\sim 2$ to 3 for follow-up imaging with the {\it HST}. This selection method is based on the successful technique used by the Sloan Lens ACS Survey (SLACS) to find over eighty-five gravitational lensed star-forming galaxies at lower redshifts \citep[e.g.][]{Bolton06, Auger09}.

This paper presents gravitational lens mass models for seventeen of the twenty-one lens candidates, as well as an analysis of the sizes and star formation rates of the reconstructed ultra-violet (UV) continuum emission from the LAE galaxies. The layout of the paper is as follows. In Section~\ref{sec:obs}, we present the high angular resolution {\it HST} observations of the rest-frame UV continuum emission from the BELLS sample of seventeen candidate LAE galaxies, from which we select the fifteen sources that we will use for our analysis. In Section~\ref{sec:models}, we describe the lens modelling procedure, which is based on an entirely Bayesian grid-based approach. In this section, we also present the recovered lens models and reconstructed sources, and we compare them with the models obtained by \citet{Shu16c}, where appropriate \citep[see also][]{Cornachione18}. In Section~\ref{sec:properties}, we investigate the intrinsic properties of the rest-frame UV continuum emission of the reconstructed sources. Finally, we compare with other samples of LAEs in the literature and discuss our results in Section~\ref{sec:disc}.

Throughout the paper, we assume $H_0=67.3\; \mathrm{km\; s^{-1}\; Mpc^{-1}}$, $\Omega_M=0.27$, $\Omega_{\Lambda}=0.73$ in a flat Universe.

\section{Data}
\label{sec:obs}

The BELLS sample was observed with the {\it HST} using the WFC3-UVIS camera and the F606W filter ($\lambda_c = 5887$ \AA; $\Delta\lambda = 2182$~\AA) between 2015 November and 2016 May (GO: 14189; PI: Bolton). In total, twenty-one candidates from the \citet{Shu16a} sample were observed for about 2600~s each. As the source redshifts are between  $z \sim 2.1$ and 2.8 and given the transmission curve of the F606W filter, these observations probe the rest-frame UV emission from young massive stars between 1250 and 2230~\AA.

\begin{figure*}
\begin{center} 
\includegraphics[width=17.5cm]{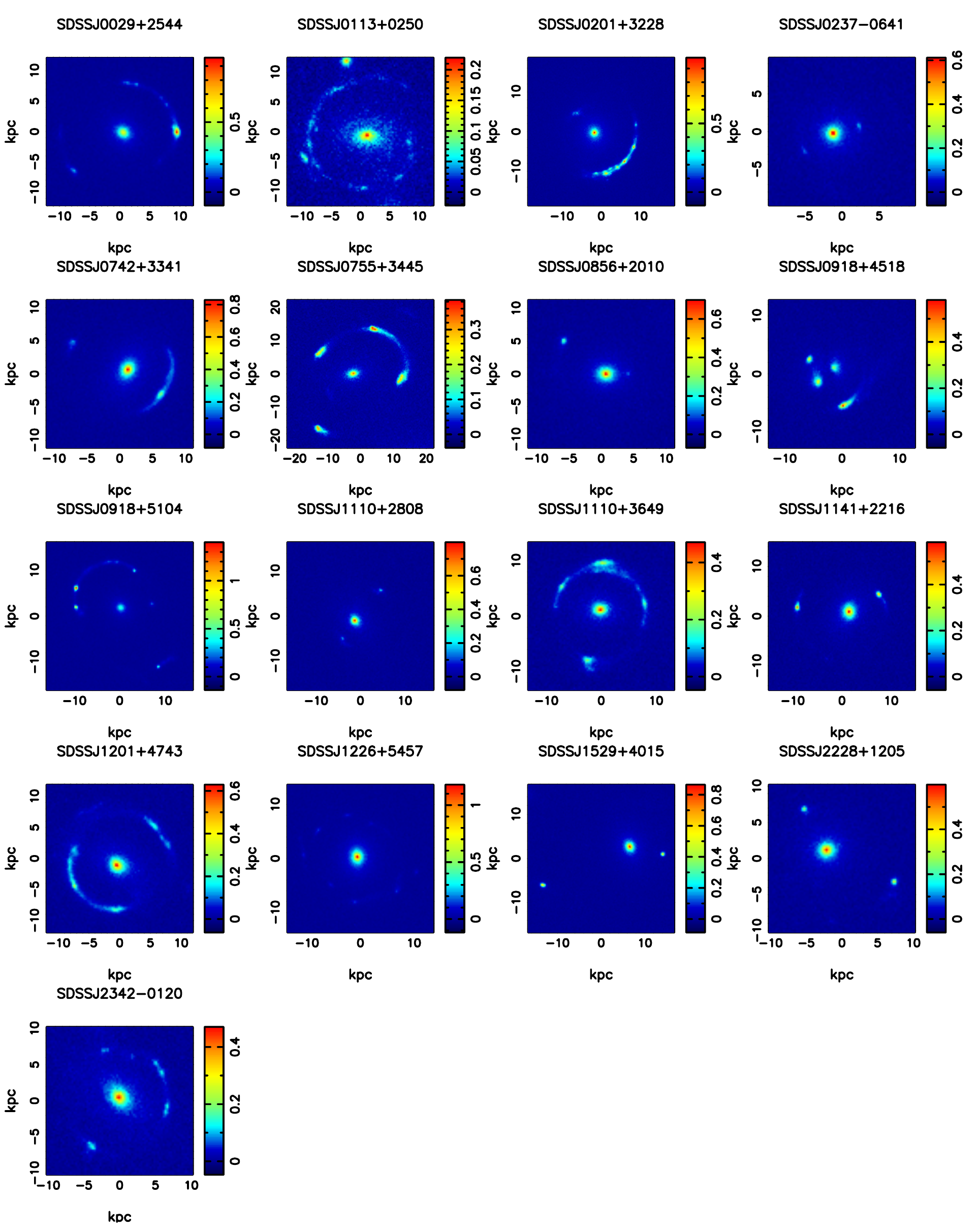}
\caption{The {\it HST} WFC3-UVIS F606W imaging of each gravitational lens system. The surface brightness scale is in electrons~s$^{-1}$. The lensing morphologies are quite varied, from nearly complete Einstein rings to very compact 2-image systems, with several examples of compound lenses (i.e. with multiple foreground galaxies causing the lensing, for example, SDSS J0918+4518).}
\label{fig:sample}
\end{center}     
 \end{figure*}

The data were retrieved from the {\it HST} archive and processed using the {\sc astrodrizzle} task that is part of the {\sc drizzlepac} package. Cut-out images for each target are shown in Fig.~\ref{fig:sample}. Out of the twenty-one candidates, three are revealed not to be strong gravitational lenses with multiple clear images of the same background galaxy: SDSS~J0054+2944, SDSS~J1116+0915 and SDSS~J1516+4954. Moreover, SDSS~J2245+0040 is also not included in our final sample due to the uncertain nature of the deflector, which has several prominent star-forming regions (the SDSS spectrum also shows [O{\sc ii}], H$\beta$, and [O{\sc iii}] emission lines). Without additional multi-band information, it is difficult to confidently distinguish structures belonging to the lens galaxy and the lensed images of the background source. Therefore, the sample used for our analysis contains seventeen gravitational lens systems \citep[see also discussion by][]{Shu16c}. The details about the individual objects in our final sample and the {\it HST} data that we have used are summarised in Table~\ref{tbl:list}.

\begin{table}
\begin{center}
\caption{Details of the gravitationally lensed LAEs used for our analysis.}
\begin{tabular}{ccccc}
\hline
Name (SDSS) & $\rm{z_{lens}}$ & $\rm{z_{src}}$ & $\lambda_{\rm rest}$ & Exp. Time\\
& & & [\AA] & [$s$]\\
 \hline
J0029$+$2544		&0.587	&2.450	&1706	&2504\\
J0113$+$0250		&0.623	&2.609	&1631	&2484\\
J0201$+$3228		&0.396	&2.821	&1540	&2520\\
J0237$-$0641		&0.486	&2.249	&1812	&2488\\
J0742$+$3341		&0.494	&2.363	&1751	&2520\\
J0755$+$3445		&0.722	&2.634	&1620	&2520\\
J0856$+$2010		&0.507	&2.233	&1821	&2496\\
J0918$+$4518		&0.581	&2.344	&1730	&2676\\
J0918$+$5104		&0.581	&2.404	&1730	&2676\\
J1110$+$2808		&0.607	&2.399	&1732	&2504\\
J1110$+$3649		&0.733	&2.502	&1682	&2540\\
J1141$+$2216		&0.586	&2.762	&1565	&2496\\
J1201$+$4743		&0.563	&2.126	&1883	&2624\\
J1226$+$5457		&0.498	&2.732	&1578	&2676\\
J1529$+$4015		&0.531	&2.792	&1553	&2580\\
J2228$+$1205		&0.530	&2.832	&1536	&2492\\
J2342$-$0120		&0.527	&2.265	&1803	&2484\\
 \hline
\end{tabular}
\label{tbl:list} 
\end{center}
\end{table}

\section{Gravitational Lens Modelling}
\label{sec:models}
Each gravitational lens system has been modelled independently with two different implementations of the Bayesian pixelated technique developed by \citet{V09}. In particular, the results presented in the following sections are based on a new version of this technique, which also fits for the light distribution from the foreground lensing galaxy. In this section we provide more details on the new features of this version, while we refer the reader to \citet{V09} and \cite{V14a} for a more detailed description of the original method, and \citet{Vegetti10a,Vegetti10b,Vegetti12} for its application to high resolution optical and infrared imaging from the {\it HST}.

\subsection{The Lens Model}
\label{ssec:modelling}

We start by considering the observed surface brightness distribution $\bmath d$ given by the combination of the lensed image $\bmath d_s$ of an unknown extended background source $\bmath s$ and the surface brightness distribution of the lensing galaxy $\bmath d_l$.  Both $\bmath d$ and  $\bmath s$ are vectors representing the surface brightness distribution on a set of pixels in the lens (i.e. observed) plane and the source plane, respectively. The grid on the lens plane is defined by the native CCD pixelation of the data, while the grid on the source plane is defined by a magnification-adapted Delaunay tessellation \citep[see][for more details]{V09}. This approach provides a pixelated surface brightness distribution for the reconstructed source that is free from any parametrised assumptions, such as S\'ersic or Gaussian light profiles, that may not adequately account for the clumpy nature of the rest-frame UV emission from the lensed sources.

We relate the relative positions of the pixels between the two planes via the lensing equation and the projected gravitational potential $\psi(x,\boldsymbol{\eta})$ of the lensing galaxy. The unknown parameters $\boldsymbol{\eta}$ defines the latter. Taking advantage of the fact that gravitational lensing conserves surface brightness and taking into account the observational noise $\bmath n$ (assumed to be Gaussian and uncorrelated among data pixels), $\bmath d$ and $\bmath s$ can be related to each other via a set of linear equations,
\begin{equation}
\mathbf{B}\left[\mathbf{L}~|~\left(\bmath{\Sigma_0}~...~\bmath{\Sigma_n}\right)~|~\bmath{1}\right]\left(\begin{array}{c}\bmath s\\ I_0\\ .\\.\\.\\I_n\\b\end{array} \right)+\bmath{n} = \bmath{d_s}+\bmath{d_l} = \bmath{d}.
\label{eq:linear_system}
\end{equation}
Here, $\mathbf{B}$ is the blurring operator that expresses the effect of the point spread function (PSF). $\bmath{\Sigma_i} I_i$ is the surface brightness distribution of the foreground gravitational lens(es). The latter is simultaneously modelled with the lens(es) mass distribution and is parametrized as elliptical S\'ersic profiles each of normalization $I_i$, effective radius $R_{e,i}$, S\'ersic index $n_{i}$ and axis ratio $q_i$, such that,
\begin{equation}
S_i\left(x,y\right) = I_i \exp \left [ - a_i \left( \left (  \frac{\sqrt{q_i^2x^2+y^2}}{R_{e,i}} \right ) ^ {1/n_i}  - 1.0 \right)  \right ] = I_i~\Sigma_i\left(x,y\right) \,,
\label{eq:S\'ersic}
\end{equation}
with $a_i = 1.9992~n_i-0.3271$. We refer to the S\'ersic parameters (excluding the linearly determined normalizations $I_i$) collectively as $\boldsymbol{\eta_l}$. The last column of the response operator given in equation (\ref{eq:linear_system}) represents a constant pedestal of amplitude $b$, expressing any residual sky background. Finally, $\mathbf{L}$ is the lensing operator and is related via the lens equation to the lens mass distribution. Here, the latter is parametrised with an external shear and an elliptical power-law profile of dimensionless surface mass density $\kappa$, given by,
\begin{equation}
\kappa\left (x,y \right ) = \frac{\kappa_0 \left ( 2 - \frac{\gamma}{2} \right ) q^{\gamma - 3/2}}{2 \left ( q^2 \left ( x^2 + r_c^2 \right ) + y^2\right )^{\left ( \gamma - 1 \right )/2}}\,, 
\label{eq:massdensity}
\end{equation}
where $\kappa_0$ is the surface mass density normalisation, $q$ is the axis ratio, $\gamma$ is the radial slope and $r_c$ the core-radius. In addition, the position angle of the elliptical mass distribution ($\theta$), the shear strength ($\Gamma$) and positional angle ($\Gamma_\theta$) are also free parameters of the model. We refer to the mass density parameters collectively as $\boldsymbol{\eta}$. The dimensionless surface mass density and the Einstein radius ($R_{\rm{ein}}$) are related to each other via
\begin{equation}
R_{\rm{ein}} = \left ( \frac{\kappa_0 \left ( 2 - \frac{\gamma}{2} \right ) q^{\left ( \gamma - 2 \right )/2}}{3 - \gamma} \right )^{1/\left ( \gamma - 1 \right )}\,.
\label{eq:einsteinradius}
\end{equation}

\subsection{Lens modelling procedure}
Redefining equation (\ref{eq:linear_system}) as
\begin{equation}
\mathbf{M}\mathbf{r} + \mathbf{n} = \mathbf{d},
\label{eq:operator} 
 \end{equation}
we derive the maximum a posteriori model parameters (MAP) with a Simulated-Annealing technique by minimizing the penalty function,
\begin{equation}
P\left(\mathbf{r}, \boldsymbol{\eta}, \mathbf{\lambda}~|~\mathbf{d}, \mathbf{H}\right) = \|\mathbf{M}\mathbf{r} - \mathbf{d}\|_2^2 + \lambda^{2} \|\mathbf{H}\mathbf{r}\|_2^2,
\label{eq:penalty} 
\end{equation}
where $\mathbf{H}$ and $\lambda$ are respectively the form and (unknown) level of regularization for the source surface-brightness distribution and are mainly a form of prior on the level of smoothness of the background galaxy \citep[see][]{Suyu06, V09}. At each step of this optimisation scheme, the corresponding most probable source $\bmath s$, sky background level $b$ and the normalisation of each S\'ersic component $I_i$ are obtained by solving the linear system
\begin{equation}
\left( \mathbf{M^T}  \mathbf{C_{d}^{-1}} \mathbf{M} + \lambda^{2}~\mathbf{H^T}\mathbf{H}\right) \mathbf{r} = \mathbf{M^T}  \mathbf{C_{d}^{-1}} \mathbf{d}.
\label{eq:source} 
\end{equation}
The main advantage of our modelling approach is that it explicitly accounts for any potential covariance in modelling of the foreground lens and background source surface-brightness; this is otherwise ignored if a model for the foreground light is estimated (e.g., by masking the pixels with significant flux from the lensed background source) and subtracted before performing the lens modelling. 

This framework will provide a more robust determination of the surface brightness distribution of the background source, but it comes at the cost of significantly increasing the dimensionality of the non-linear parameter space, which is comprised of the foreground lens surface brightness and mass parameters, $\boldsymbol{\eta_l}$ and $\boldsymbol{\eta}$. To overcome this difficulty, the modelling procedure is performed in three steps. First we mask out the emission from the lens galaxy and, given the observed lensed surface brightness distribution $\bmath d$, we optimize for the lens mass parameters $\boldsymbol{\eta} = \{\kappa_0,\theta,q,x,y,\gamma,\Gamma,\Gamma_{\theta}\}$ and the source regularization level.  Then, using this as a starting point, we parameterise the surface brightness distribution of the lens galaxy as a sum of multiple elliptical S\'ersic profiles and optimise for the corresponding parameters, $\boldsymbol{\eta_l}$. In the third and final step, we optimise simultaneously for the mass and the light distribution of the deflector and the source regularisation level.

Each lens system is modelled several times, allowing for different forms of regularisation (i.e. gradient or curvature) and different source plane grids. In the next section, we report the models with the highest Bayesian evidence. To quantify the uncertainties and probe any degeneracies among the model parameters, we use {\sc MultiNest} \citep{Multinest13} to explore the parameter space within the prior volume. The latter is defined by uniform prior distributions on all parameters except for the source regularisation constant, for which we adopt a uniform prior in logarithmic space. 

\subsection{Lens modelling results}
\label{ssec:results}

We present the inferred lens models and reconstructed sources in Fig.~\ref{fig:res_1}. The lens model parameters for each system are listed in Table~\ref{tbl:lenspar} and the parameters for the S\'ersic surface brightness distributions of the lensing galaxies are reported in Table~\ref{tbl:sersic}. 

The same systems have also been modelled by \cite{Shu16c} (referred to as S16 in the following) under the assumption of a singular isothermal ellipsoid (SIE) mass distribution plus external shear. While our models allow for a varying slope, we don't find a significant departure from $\gamma =2$. In general, our lens mass parameters are in a reasonable agreement with those reported by S16, with differences at the 10 percent level at most, except for SDSS~J0201+3228, SDSS~J0755+3445 and SDSS~J0918+5104. We discuss these discrepancies in more detail below. In three other cases, SDSS~J0237$-$0641, SDSS~J1226+5457 and SDSS~J2342$-$0120, the lens position angles differ by more than 10 percent but, since all these systems are well fitted by a circular mass distribution with axis ratios close to unity, this is just an apparent discrepancy.

We also find an agreement at the 4 percent level on the lens light parameters, except for the effective radii. This discrepancy is possibly related to a degeneracy between the effective radius and the S\'ersic index, for which S16 do not report any values. We find that for SDSS~J0237$-$0641 and SDSS~J1226+5457 the light position angles differ by more than 4 percent, a discrepancy which is again explained by the roundness of the light profile.

From a qualitative comparison of the reconstructed sources, we notice that S16 report generally smoother and less compact sources than our reconstructions. This difference could be related to different choices of regularisation form and level. However, due to the lack of detailed information about the source regularisation in S16, it is not possible to draw strong conclusions on the origin of such a discrepancy, with the choice of masking also potentially playing an important role. In particular, six of the reconstructed sources are qualitatively different from the ones reported in S16, and a more detailed discussion of these discrepancies is given in the following subsections. We note that our two independent lensing codes lead to reconstructed sources that are consistent with each other.

In general, our models provide a better match, expressed in terms of image residuals, to the data. This result is particularly important to avoid the false detection of mass substructures in the lens galaxies or sub-haloes along the line-of-sight. Nevertheless, in some cases, our source models do not fully recover the data for pixels with a surface brightness much higher than that of neighbouring pixels. We do not believe that this is due to poor quality lens mass models, but is related instead to the regularisation imposed on the reconstructed source. Due to the pixelated nature of our source reconstruction technique, a prior on the smoothness of the source has to be imposed via equation (\ref{eq:penalty}). In practice, this is done by minimising either the gradient or the curvature of the surface brightness between neighbouring pixels on the source plane.  While this allows us to avoid noise fitting and by-pass the ill-defined character of equation (\ref{eq:linear_system}), it also makes it extremely challenging to recover the brightness of the brightest pixels where an extremely high dynamic range characterises the lensed images.

From the parameter space exploration, we find mean values that are consistent with our MAP parameters up to at most the 20 per cent level (see Tables~\ref{tbl:MCMClens} and \ref{tbl:MCMCser}). This difference often corresponds to a several $\sigma$ discrepancy, which is related to the fact that the quoted errors are purely statistical and do not include the systematic uncertainty. We have estimated the latter by comparing the results obtained with the two different Bayesian techniques used for the modelling procedure and the results by \cite{Shu16c}, and found them to be at the 10 per cent level.

\begin{figure*}
\begin{center} 
\includegraphics[width= 17.7cm, keepaspectratio]{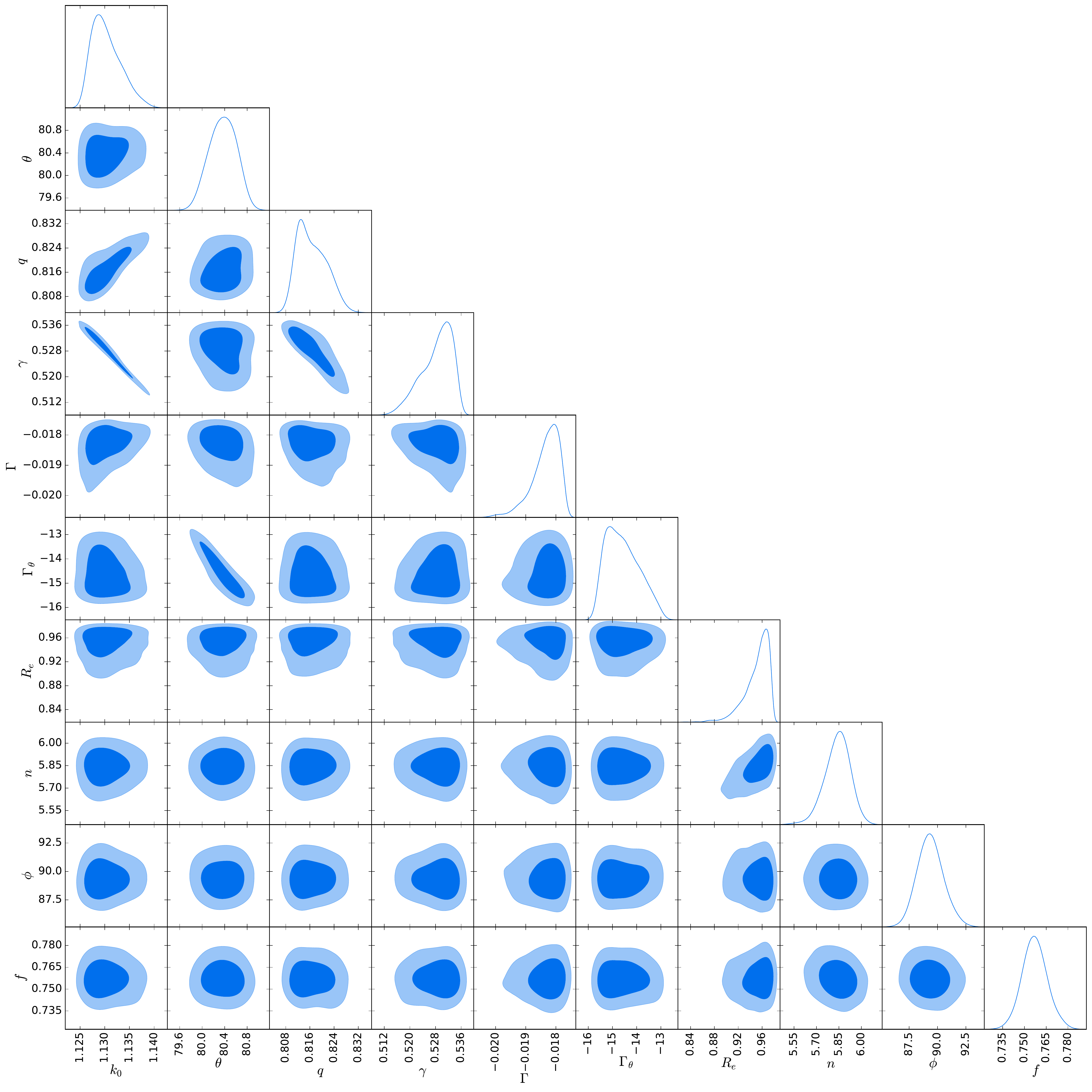}
\caption{The posterior probability distribution for the mass and light parameters of the lens galaxy in SDSS~J1110+3649. The contours for the 68\% and 95\% confidence level are shown. Apart from the usual degeneracies among the lens mass parameters, no degeneracy is found between the mass and light models.}
\label{fig:nodeg}
\end{center}     
\end{figure*}

Apart from the usual degeneracies among some of the mass lens parameters (e.g. $\kappa_0$ and $\gamma$), we find no covariance between the lens light and mass parameters (e.g. see Fig.~\ref{fig:nodeg} for a typical case). This result is important as it implies that our mass models are robust against changes in the light model. This result is explicitly tested by re-modelling five systems after the lensing galaxy light has been fitted with a B-spline technique and subtracted from the data, inferring lens mass parameters consistent with our MAP parameters at the 10 percent level (see also Section~\ref{ssec:bspline}). However, by comparing our results with those from S16, we find a certain level of degeneracy between the lens light parameters and the reconstructed sources in a way that also depends on the specific choice of masking and regularisation.

\begin{table*}
\caption{The maximum a posteriori model parameters (MAP) for the gravitational lens mass models. The parameters of the mass density distribution reported are as follows: $\kappa_0$ is the normalization, $\rm{\theta}$ is the position angle with respect to north, $\rm{\gamma}$ is the radial slope and $q$ is the axis ratio. $\rm{\Gamma}$ and $\rm{\Gamma_{\theta}}$ are respectively the magnitude and the position angle of the external shear (with respect to north).}
\begin{tabular}{ccccccc}
\hline
Name (SDSS) & $\kappa_0$ & $\rm{\theta}$ & $q$ & $\rm{\gamma}$ & $\rm{\Gamma}$ & $\rm{\Gamma_{\theta}}$\\
& [arcsec] & [deg] &  &  &  & [deg] \\
 \hline
J0029$+$2544		&1.295 &34.237   &0.823 &2.101	 &0.045	 &28.935\\
J0113$+$0250		&1.226 &81.671   &0.730 &2.046   &0.001	 &184.002\\
				&0.065 &218.768  &0.966 &2.000   & &  \\
                        		&0.172 &237.445  &0.922	&2.000   & &  \\
J0201$+$3228		&1.650 &52.764   &0.896	&1.959   &0.049  &32.514 \\
J0237$-$0641		&0.687 &39.575   &0.881	&1.914   &0.008  &237.567\\
J0742$+$3341		&1.197 &152.279  &0.824	&2.026   &0.021  &353.566\\
J0755$+$3445		&1.926 &105.347  &0.533	&1.898   &0.230  &31.865\\
J0856$+$2010		&0.960 &121.504  &0.653	&2.024   &0.073  &76.157	\\
J0918$+$4518		&0.444 &41.410   &0.843 &2.036   &0.086  &82.185	\\
				&0.409 &29.843   &0.926	&2.015   & &  \\
J0918$+$5104		&1.600 &16.995   &0.682	&2.143   &0.253  &122.144\\
J1110$+$2808		&0.992 &41.672   &0.892 &2.016   &0.020  &317.727 \\
J1110$+$3649		&1.152 &79.922   &0.865	&1.974   &0.020  &255.758\\
J1141$+$2216		&1.281 &150.356  &0.751	&1.983   &0.003  &14.071	\\
J1201$+$4743		&1.139 &38.496   &0.780	&2.149   &0.008  &49.842\\
J1226$+$5457		&1.351 &72.448   &0.970	&2.055   &0.153  &161.277\\
J1529$+$4015		&2.233 &41.124   &0.542	&1.997   &0.007  &230.436\\
J2228$+$1205		&1.290 &176.851  &0.923	&1.966   &0.034  &3.539  \\
J2342$-$0120		&1.033 &43.019   &0.798	&2.159   &0.021  &272.015 \\
 \hline
\end{tabular}
\label{tbl:lenspar} 
\end{table*}

\begin{table*}
\caption{The maximum a posteriori model parameters (MAP) for the S\'ersic fits to the light of the lensing galaxies. For each system, we report for each lens and component (where multiple light profiles are fitted) the effective radius $R\rm{_e}$, the S\'ersic index $n$, the position angle $\phi$ (with respect to north) and the axis ratio $f$.}
\begin{tabular}{cccccccc}
\hline
Name (SDSS)     &lens  &component  &R$\rm{_e}$    & $n$    &$\rm{\phi}$   & $f$ \\
			    &      &           &[arcsec] 	  &     &[deg]  &  \\
 \hline
J0029$+$2544 	&   &   &0.759  &4.531  &51.599  &0.800\\
J0113$+$0250 	& 1 &   &1.277  &3.472  &85.330 &0.583\\
				  			& 2 &   &0.103  &2.558  &120.486 &0.949\\
				 			& 3 &   &0.292  &2.574  &206.409 &0.579\\
J0201$+$3228 	&   &I  &2.967  &2.844  &25.814  &0.890\\
				 			&   &II &0.241  &5.557  &149.106 &0.915\\
J0237$-$0641 	&   &   &4.188  &8.593  &3.293   &0.996\\
J0742$+$3341 	&   &   &2.057  &6.356  &149.261 &0.708\\
J0755$+$3445 	&   &   &0.581  &4.055  &103.288 &0.620\\
J0856$+$2010 	&   &   &0.950  &4.752  &94.637  &0.784\\
J0918$+$4518 	& 1 &I  &0.640  &4.135  &163.164 &0.568\\
				  			&   &II &0.059  &3.871  &176.439  &0.630\\
				  			& 2 &   &1.165  &4.234  &40.174  &0.901\\
J0918$+$5104 	&   &I  &0.485  &2.817  &42.549  &0.878\\
				  			&   &II &0.065  &2.339  &19.194  &0.851\\
J1110$+$2808 	&   &I  &0.675  &3.468  &56.278  &0.780\\ 
				 			&   &II &0.609  &3.591  &24.236  &0.800\\
J1110$+$3649 	&   &   &0.889  &5.625  &88.740  &0.756\\
J1141$+$2216 	&   &I  &0.608  &3.621  &153.127 &0.818\\
				  &   &II &0.249  &3.640  &151.045 &0.822\\
J1201$+$4743 	&   &I  &1.476  &5.174  &62.968  &0.761\\
				  &   &II &1.132  &5.336  &52.551  &0.750\\
J1226$+$5457 	&   &   &0.708  &3.920  &182.989 &0.823\\
J1529$+$4015 	&   &   &1.630  &5.945  &19.098  &0.805\\
J2228$+$1205 	&   &   &0.779  &4.760  &102.483 &0.933\\
J2342$-$0120 	&   &   &2.123  &5.903  &43.973  &0.655\\
\hline
\end{tabular}
\label{tbl:sersic} 
\end{table*}

\begin{table*}
\caption{Mean values and relative errors for the lens mass models, derived using {\sc MultiNest}. The uncertainties are purely statistical and do not include systematic errors. Note that for the two additional lenses in SDSS~J0113$+$0250, the radial density slope was kept fixed at a value of $\gamma = 2$ (corresponding to a SIE) and therefore they are not reported here.}
\begin{tabular}{ccccccc}
\hline
Name (SDSS)   &$\kappa\rm{_0}$   &$\rm{\theta}$    & $q$    &$\rm{\gamma}$    &$\rm{\Gamma}$    &$\rm{\Gamma_{\theta}}$ \\
			  &[arcsec] 	&[deg] 	  &     &                 &                 &[deg]           \\
 \hline
J0029$+$2544	&1.320 $\pm$ 0.003  &35 $\pm$ 4   &0.94 $\pm$ 0.02   &1.886 $\pm$ 0.002   &0.05 $\pm$ 0.002   &28 $\pm$ 2\\
J0113$+$0250  	&1.219 $\pm$ 0.005  &80.6 $\pm$ 0.3   &0.72 $\pm$ 0.01   &2.08 $\pm$ 0.01	 &0.001 $\pm$ 0.0001 &165 $\pm$ 14\\
                   	&0.058 $\pm$ 0.003  &236 $\pm$ 15 &0.86 $\pm$ 0.05   &	&	&\\
                   	&0.176 $\pm$ 0.005  &226 $\pm$ 9  &0.82 $\pm$ 0.04   &	&	&\\
J0201$+$3228   	&1.6604 $\pm$ 0.0001 &53.5 $\pm$ 0.1   &0.9105 $\pm$ 0.0005  &1.8638 $\pm$ 0.0004	 &0.0489 $\pm$ 0.0002	&34.91 $\pm$ 0.02\\
J0237$-$0641	&0.7 $\pm$ 0.04   &36 $\pm$ 0.9   &0.8 $\pm$ 0.02    &1.860 $\pm$ 0.04   &0.007 $\pm$ 0.001   &237.1 $\pm$ 3.3\\
J0742$+$3341	&1.208 $\pm$ 0.005  &151.9 $\pm$ 0.4  &0.835$\pm$ 0.007   &1.98 $\pm$ 0.01   &0.021 $\pm$ 0.001	  &352.6 $\pm$ 0.4\\
J0755$+$3445	&1.86 $\pm$ 0.01  &104.94 $\pm$ 0.05  &0.531 $\pm$ 0.002   &1.834 $\pm$ 0.01   &0.232 $\pm$ 0.0005	&30.18 $\pm$ 0.1\\
J0856$+$2010	&1.093 $\pm$ 0.0004 &100 $\pm$ 0.6  &0.775 $\pm$ 0.001   &1.820 $\pm$ 0.0004	 &0.072 $\pm$ 0.001	 &84.8 $\pm$ 0.7\\
J0918$+$4518	&0.362 $\pm$ 0.005  &47 $\pm$ 2   &0.84 $\pm$ 0.03   &2.15 $\pm$ 0.01	 &0.093 $\pm$ 0.003	  &78 $\pm$ 2\\
					  		      &0.48 $\pm$ 0.01  &31 $\pm$ 3   &0.938 $\pm$ 0.03   &2.01 $\pm$ 0.01	 &	&\\
J0918$+$5104	&1.560 $\pm$ 0.001  &20.1 $\pm$ 0.1   &0.703 $\pm$ 0.001   &2.254 $\pm$ 0.002   &0.267 $\pm$ 0.001	 &124.2 $\pm$ 0.03\\
J1110$+$2808	&0.882 $\pm$ 0.004  &45 $\pm$ 1   &0.847 $\pm$ 0.007   &2.210 $\pm$ 0.003   &0.020 $\pm$ 0.002	 &320 $\pm$ 3\\
J1110$+$3649	&1.116 $\pm$ 0.007  &80.4 $\pm$ 0.3   &0.79 $\pm$ 0.01    &2.10 $\pm$ 0.01	 &0.016 $\pm$ 0.001	 &254 $\pm$ 1\\
J1141$+$2216	&1.269 $\pm$ 0.004   &155.8 $\pm$ 0.3  &0.76 $\pm$ 0.01   &1.92 $\pm$ 0.01	 &0.0034 $\pm$ 0.0002	&14.1 $\pm$ 0.7\\
J1201$+$4743	&1.147 $\pm$ 0.004  &39.1 $\pm$ 0.3   &0.799 $\pm$ 0.004   &2.112 $\pm$ 0.006	 &0.010 $\pm$ 0.0005	&42 $\pm$ 2\\
J1226$+$5457	&1.355 $\pm$ 0.004  &62.3 $\pm$ 0.4   &0.915 $\pm$ 0.003   &2.04 $\pm$ 0.02	 &0.162 $\pm$ 0.002	 &159.9 $\pm$ 0.1\\
J1529$+$4015	&2.2 $\pm$ 0.01    &37.1 $\pm$ 0.2   &0.504 $\pm$ 0.01   &2.00 $\pm$ 0.01  & 0.0075 $\pm$ 0.0004 & 222$\pm$ 3\\
J2228$+$1205	&1.266 $\pm$ 0.004  &176.9 $\pm$ 0.2  &0.899 $\pm$ 0.005   &2.04 $\pm$ 0.01	&0.033 $\pm$ 0.001	   &3.6 $\pm$ 0.2\\
J2342$-$0120	&1.003 $\pm$ 0.001 &40.8 $\pm$ 0.1   &0.8765 $\pm$ 0.0004  &2.0596 $\pm$ 0.001	&0.0196 $\pm$ 0.0001	   &272.13 $\pm$ 0.05\\
 \hline
\end{tabular}
\label{tbl:MCMClens} 
\end{table*}

\begin{table*}
\caption{Mean values and relative errors for the lens light derived with {\sc MultiNest}. The uncertainties are purely statistical and do not include systematic errors.}
\begin{tabular}{ccccccc}
\hline
Name (SDSS)     &lens  &component  &R$\rm{_e}$    & $n$    &$\rm{\phi}$   & $f$ \\
			    &      &           &[arcsec] 	  &     &[deg]  &  \\
 \hline
J0029$+$2544	&  &   &0.74 $\pm$ 0.07    &5 $\pm$ 0.3   &51 $\pm$ 6   &0.8 $\pm$ 0.1\\
J0113$+$0250	&1 &   &1.50 $\pm$ 0.03      &3.70 $\pm$ 0.04   &83.5 $\pm$ 0.6      &0.59 $\pm$ 0.01\\
                    	&2 &   &0.102 $\pm$ 0.005    &2.4 $\pm$ 0.2     &125 $\pm$ 10         &0.95 $\pm$ 0.02\\
                    	&3 &   &0.347 $\pm$ 0.003    &2.08 $\pm$ 0.01   &246.8 $\pm$ 0.7      &0.49 $\pm$ 0.02\\
J0201$+$3228	&  &I      &2.79 $\pm$ 0.03    &2.61 $\pm$ 0.02   &27.56 $\pm$ 0.4   &0.862 $\pm$ 0.004\\
        			&       &II     &0.240 $\pm$ 0.004     &5.49 $\pm$ 0.07   &151.1 $\pm$ 1.6  &0.88 $\pm$ 0.01\\
J0237$-$0641	&  &	&4.7 $\pm$ 0.2    &8.8 $\pm$ 0.2   &3.3 $\pm$ 0.4 &0.984 $\pm$ 0.008\\
J0742$+$3341	 &  &   &2.21 $\pm$ 0.04    &6.5 $\pm$ 0.1   &147.6 $\pm$ 0.5  &0.705 $\pm$ 0.004\\
J0755$+$3445	&  &	&0.62 $\pm$ 0.01    &4.26 $\pm$ 0.05   &102.012 $\pm$ 0.6  &0.617 $\pm$ 0.006\\
J0856$+$2010	&  &	&0.81 $\pm$ 0.02    &4.44 $\pm$ 0.08   &95.4 $\pm$ 0.6   &0.779 $\pm$ 0.004\\
J0918$+$4518	&1 &I  &0.56 $\pm$  0.03   &3.42 $\pm$ 0.08   &164 $\pm$ 2  & 0.58 $\pm$ 0.03\\
								  &  &II &0.055 $\pm$ 0.002    &3.14 $\pm$ 0.03   &176.4 $\pm$ 0.4  &0.61 $\pm$ 0.02\\
           					      &2 &   &1.34 $\pm$ 0.04    &4.5 $\pm$ 0.1   &40 $\pm$ 4   &0.89 $\pm$ 0.02\\
J0918$+$5104	&  &I	&0.50 $\pm$ 0.01    &2.90 $\pm$ 0.05   &40 $\pm$ 2   &0.882 $\pm$ 0.006\\
								  &  &II	&0.071 $\pm$ 0.003    &2.6 $\pm$ 0.1   &19 $\pm$ 3   &0.92 $\pm$ 0.04\\
J1110$+$2808	&  &I	&0.78 $\pm$ 0.02    &4.03 $\pm$ 0.09   &48 $\pm$ 2   &0.65 $\pm$ 0.02\\
								  &	&II	&0.70 $\pm$ 0.02    &3.73 $\pm$ 0.04   &21 $\pm$ 1   &0.83 $\pm$ 0.01\\
J1110$+$3649	&  &	&1.04 $\pm$ 0.02    &6.05 $\pm$ 0.08   &89 $\pm$ 1   &0.756 $\pm$ 0.008\\
J1141$+$2216	&  &I	    &0.65 $\pm$ 0.01    &3.8 $\pm$ 0.05   &153 $\pm$ 1  &0.813 $\pm$ 0.007\\
                			&	&II 	&0.25 $\pm$ 0.01    &3.7 $\pm$ 0.2    &150 $\pm$ 2  &0.79 $\pm$ 0.03\\
J1201$+$4743	&  &I	&1.72 $\pm$ 0.04    &4.18 $\pm$ 0.03   &62 $\pm$ 3   &0.618 $\pm$ 0.007\\
								  &  &II	&0.94 $\pm$ 0.03    &5.34 $\pm$ 0.08   &47 $\pm$ 1   &0.737 $\pm$ 0.006\\
J1226$+$5457	&  &	&0.69 $\pm$ 0.02    &3.87 $\pm$ 0.06   &181.3 $\pm$ 0.6  &0.821 $\pm$ 0.004\\
J1529$+$4015	&  &	&1.68 $\pm$ 0.05    &6 $\pm$ 0.1     &19.4 $\pm$ 0.5     & 0.805 $\pm$ 0.005\\
J2228$+$1205	&  &	&0.83 $\pm$ 0.02    &5.16 $\pm$ 0.06   &102 $\pm$ 4  &0.93 $\pm$ 0.01\\
J2342$-$0120	&  &	&2.16 $\pm$ 0.03    &6.02 $\pm$ 0.05   &41.9 $\pm$ 0.5   &0.652 $\pm$ 0.005\\
 \hline
\end{tabular}
\label{tbl:MCMCser} 
\end{table*}

\subsection{Specific cases}
\label{ssec:specific_cases}

Here, we give comments on a few special systems and on those systems for which our results are significantly different from those obtained by S16.

\subsubsection{SDSS~J0029+2544}

A particularly intense peak of surface brightness is visible in the images in the arc directly west from the lensing galaxy; this does not have a corresponding high-surface-brightness peak in the counter-image and is only partially recovered in our lens model. Multi-band observations would be required to identify its origin. 
While the image residuals and the mass parameters are in agreement with those reported by S16, our source has a different structure. In particular, 
our source is composed of one main component located on the cusp of the caustic and a second and fainter component lying on its left while in the source reconstructed by S16 the second component is much brighter. We believe the origin of this discrepancy to lie on the specific choice of masking and related regularisation effects. Finally, we infer light parameters that, except for the S\'ersic effective radius (see discussion above), are in agreement with those reported by S16.

\subsubsection{SDSS~J0113+0250}

This system is composed of three lenses, with the two secondary ones bending the arc respectively in the north and southeast direction. The third lens is located at a projected distance of approximately 2 arcsec from the main lens. For all three deflectors, the brightness profile is fitted using only one S\'ersic component. The source has a particularly clumpy profile. Unlike S16, we find a significant improvement in terms of the Bayesian evidence and image residuals when the third lens is added to the mass distribution.

\subsubsection{SDSS~J0201+3228}

Unlike S16, we infer a lens mass profile that is round, $q=0.9$, rather than elliptical $q=0.7$. For comparison, we have remodelled this dataset with an SIE profile of fixed axis ratio 0.7 and found that the rounder model is preferred by the Bayesian evidence, with a $\Delta\log E \simeq$ 900.
It should be stressed that both of our independent models based on our two codes disfavour an elliptical mass profile. Moreover, neither the surface brightness profile of the lens nor the lensing configuration from Fig.~\ref{fig:sample} suggests that this is an unusually elliptical lens. It may be that the difference in ellipticity between the S16 model and ours is counteracted by a difference in the external shear strength. However, we do not know with certainty, since S16 do not report the shear value for this lens. Finally, the lens galaxy light of this system was fitted using a sum of two S\'ersic profiles.

The image-plane residuals for SDSS~J0201+3228 are probably the worst case within the sample, with 4 knots of positive and negative emission that have normalised residuals around $2.5\sigma$ along the extended arc to the south-west. We note that the residuals in S16 are significantly worse for both their S\'ersic and grid-based source models. The strong residuals may be due to either the complex source structure requiring an adaptive regularisation to account for both the smooth and compact structure (requiring a higher dynamic range) or a mismatch between the elliptical power-law model and the actual mass distribution of the deflector. In a companion paper, we investigate more complex lens models for this system using grid-based corrections to the lensing potential, but we find that these do not improve the image residuals significantly. Therefore, any inference of the source properties for this object is left for a future analysis when multi-band data are available.

\subsubsection{SDSS~J0237$-$0641, SDSS~J0856+2010, SDSS~J1141+2216}

For these three systems, our reconstructed sources appear significantly different from those presented by S16, both regarding their position with respect to the caustic and their shape. The models for SDSS~J0237$-$0641 are comparable in terms of residuals, while our models for SDSS~J0856+2010 and SDSS~J1141+2216 seem to fit the data better as they present less significant residuals. The mass and light models are consistent for the three systems, except for the effective radii of the S\'ersic model. For the system SDSS~J1141+2216, 
a direct comparison with the S\'ersic fitting results by S16 is not possible as we have parametrised the lens light distribution with two components rather than one. 
We believe the discrepancy among the sources to be related to the proximity of the lensing galaxies to the lensed emission, especially in SDSS~J0237$-$0641. This configuration makes it harder to distinguish between lensed and lens emission, and effectively introduces a degeneracy between the light model of the foreground lens and the surface brightness of the reconstructed source, which depends on the specific choice of masking. It should also be noted that these are all doubly-imaged systems; therefore the data provides us with relatively limited information to constrain the mass and the source model.

\subsubsection{SDSS~J0755+3445}

The modelling of this quadruple image system was possible only through allowing for a strong external shear. In fact, this can already be inferred by noting that the image separation between the central part of the arc and the counter-image is significantly larger than the aperture of the arc, denoting the presence of an external element that perturbs the lensing configuration. Another component of the source is lensed in the two opposite and faint images located along the southwest-northeast diagonal. The galaxy to the southwest direction was found not to have any influence on the lens mass model and was therefore ignored in the modelling procedure. 

\begin{figure}
\begin{center}
\includegraphics[height=6cm,keepaspectratio]{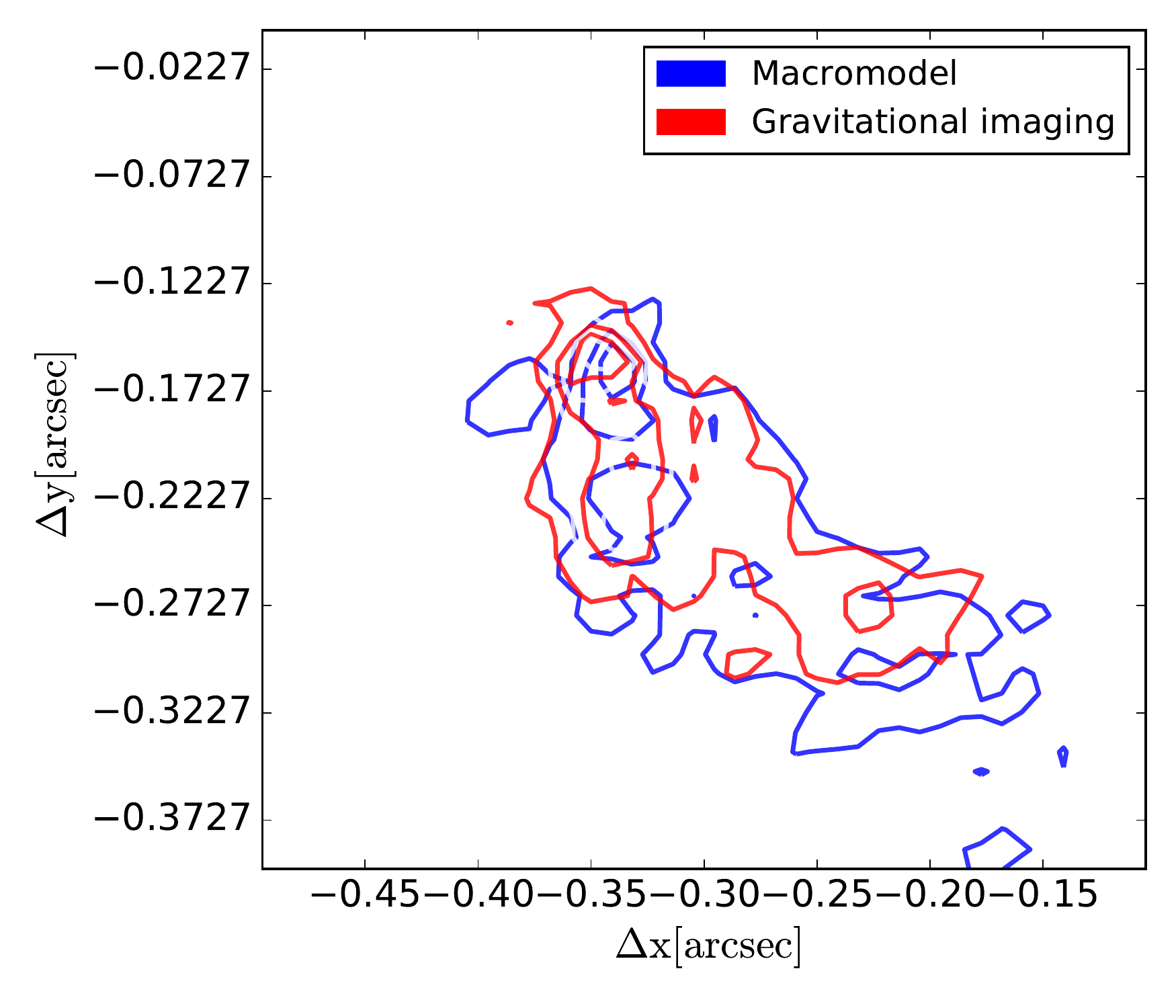}
\caption{Reconstructed source structure for the lens system {\rm{SDSS~J0755+3445}}. The blue contours show the source structure inferred under the assumption of a single power-law mass model for the lens galaxy. The red contours show the reconstructed source when the system is modelled with the gravitational imaging technique.}
\label{fig:contdata}
\end{center} 
\end{figure}

The LAE source of this system was also reconstructed via the \textit{gravitational imaging technique} (see VK09). We refer to our follow-up paper for a more detailed description of this procedure. We find that the diffused low-level pixellated convergence corrections which are added to the single power-law model result in a more compact source with a less pronounced tail as shown in Figure~\ref{fig:contdata}. We therefore conclude that a single power-law model is not representative of the mass distribution of the deflector and any inference of the source properties is left for a future analysis when multi-band data are available.

\subsubsection{SDSS~J0918+4518}
Both the light and mass models of the lens are in disagreement with those reported by S16. 
The structure of the source is also significantly different, as expected from the divergent lens models. In particular, according to our model, both deflectors have approximately equal lensing strength and are both almost round with an axis ratio of $q\simeq 0.9$. Instead, one of the lenses in the model by S16 is strongly elliptical, with an axis ratio of $q=0.132$, and has a different lensing strength. When using the lens parameters by S16 to model this system, we find that our model is preferred with a difference in Bayesian evidence of $\Delta\log E \simeq$ 200. Moreover, our residuals are less prominent, and our results are consistent between our two independent lens modelling codes.

\subsubsection{SDSS~J0918+5104}

We find that a large external shear had to be allowed to account for the general lensing configuration of this system and the displacement of the counter-image towards the southwest direction. We find that the foreground object to the west of the primary lens is not contributing significantly to the lensing signal, and we did not include its contribution into the final mass model. However, we had to take into account for its brightness distribution which was modelled with two S\'ersic profiles, while only one component was necessary to account for the lens galaxy emission.
Our mass model requires a stronger ellipticity and a stronger external shear in comparison with the one presented by S16. We remodelled this dataset with an SIE profile of fixed axis ratio $q=0.985$ and fixed shear strength $\rm{\Gamma} = 0.18$, as determined by S16, and found that our model is preferred by the Bayesian evidence, with a $\Delta\log E \simeq 300$. Finally, we remark that our reconstructed source appears more compact and closer to the caustic. This result is probably due to the difference between the two mass models.

\subsubsection{SDSS~J1110+2808}
The image residuals and the appearance of the source are consistent with the results presented by S16, and the lens models are in agreement at the 3~percent level. However, unlike S16, we find that the lens light profile is better fitted with two rather than one S\'ersic component. 

\subsubsection{SDSS~J1529+4015}

This system is composed of two sources at two different redshifts, as can be seen from the different image separation of the two sets of arcs in Fig.~\ref{fig:sample}. The first source corresponds to the two bright images on the opposite sides of the lens galaxy, and the other source is lensed into the faint arc on the right with the counter-image located below the bright image on the left. Since the second set of images is too faint to be used to reconstruct the lens model it is masked out of the data.

\subsection{B-spline subtracted data}
\label{ssec:bspline}

We find that for five systems, namely SDSS~J0113+0250, SDSS~J0918+4518, SDSS~J1110+2808, SDSS~J1141+2216 and SDSS~J2342$-$0120, a simple S\'ersic profile (with one or more components) does not provide a good fit to the complex light structure of the lens galaxy. However, as we are mainly interested in the lens mass parameters and the background source properties, we test their robustness by modelling a version of the data where the lens galaxy light has been previously fitted and subtracted using a B-spline method. In agreement with the results from {\sc MultiNest}, we find no covariance between the mass and light parameters, and we recover lens mass parameters that are in agreement with those derived from the simultaneous fit of the lens mass and light. Moreover, since for these systems the S\'ersic profiles provide a bad fit only to the central regions of the lens galaxy light, we find that also the reconstructed sources are consistent among the two different modelling approaches.

\subsection{Testing for systematics in the source structure}

Finally, we have quantified how the imposed regularisation can affect the compactness and morphology of the reconstructed source and our results. For this, we have performed a series of simulations by creating mock Gaussian sources of different sizes and lensing them through the best-fitting lens model for SDSS~J0029+2544. These are then convolved with the {\it HST} PSF to create synthetic datasets, and reconstructed using the modelling procedure described in Section~\ref{ssec:modelling}. We then compare the size of the original Gaussian source with the reconstructed source that includes the effects of the regularisation. In all cases, we find that we robustly recover the properties of sources as small as 30~mas.

\begin{figure*}
\begin{center} 
\stackon[5pt]{\includegraphics[width= 17.5cm, viewport=29 50 565 186, clip]{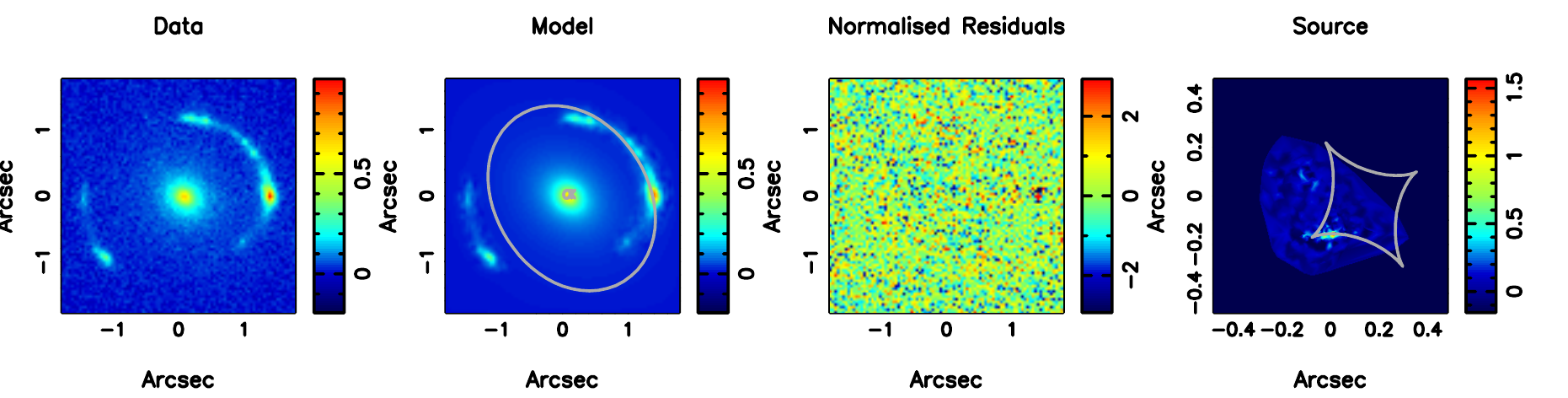}}{SDSS~J0029+2544}
\stackon[5pt]{\includegraphics[width= 17.5cm, viewport=29 50 565 186, clip]{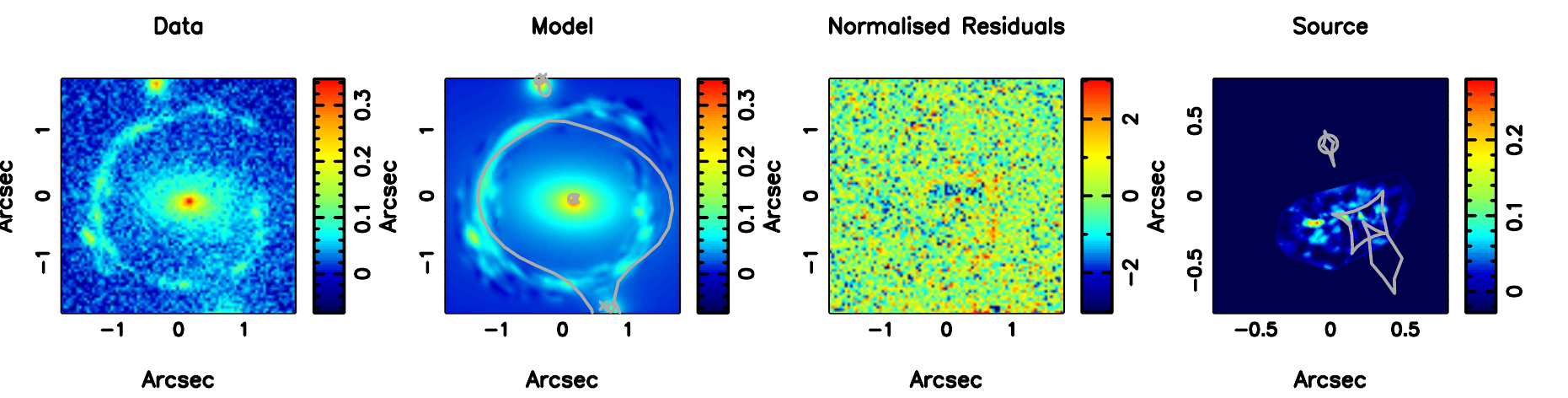}}{SDSS~J0113+0250}
\stackon[5pt]{\includegraphics[width= 17.5cm, viewport=29 50 565 186, clip]{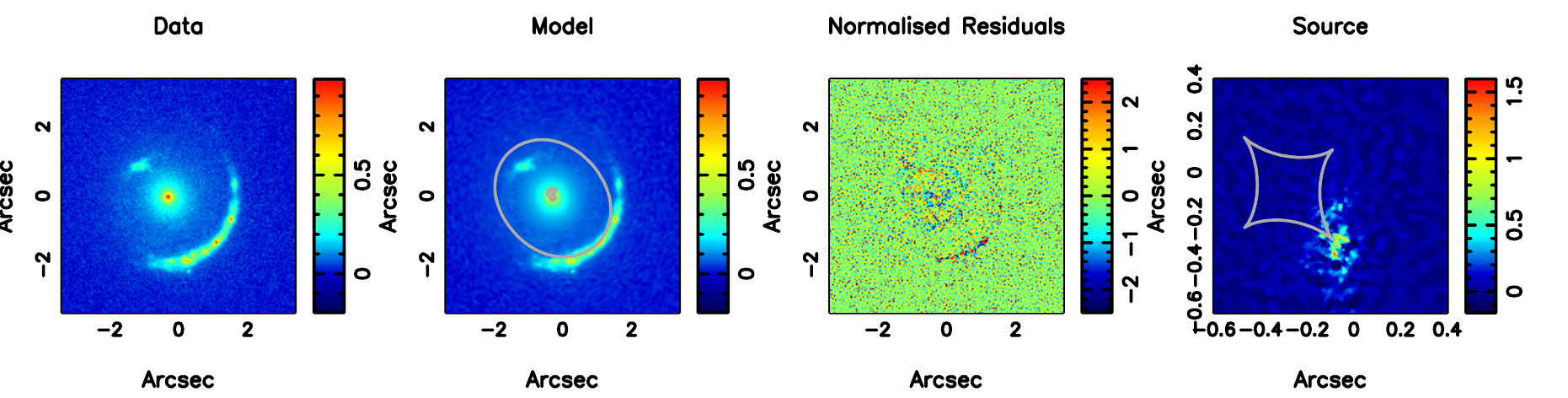}}{SDSS~J0201+3228}
\stackon[5pt]{\includegraphics[width= 17.5cm, viewport=29 50 565 186, clip]{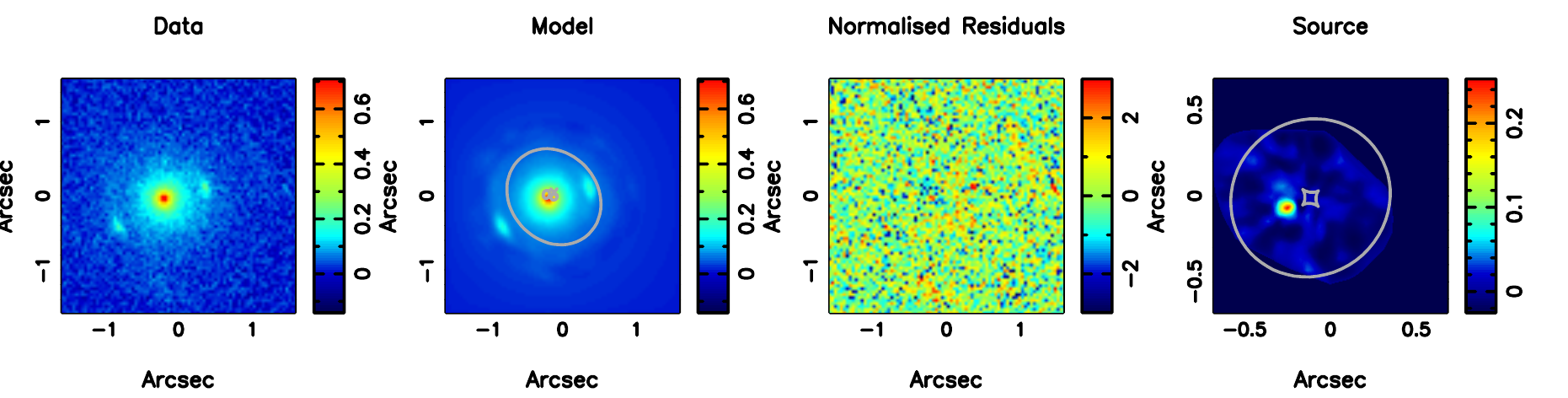}}{SDSS~J0237$-$0641}
\caption{Models for the gravitational lens systems, given in the same order as in Table~\ref{tbl:lenspar}. Each row shows (left) the input {\it HST} F606W imaging, (middle-left) the reconstructed model for lens-plane surface brightness distribution of the gravitational lensing galaxy and the gravitationally lensed LAE galaxy, (middle-right) the normalized image residuals in units of $\sigma$, and (right) the reconstructed surface brightness distribution of the LAE galaxy. Shown in grey are the gravitational lens mass model critical curves in the lens-plane and the caustics in the source-plane. \label{fig:res_1}}
\end{center}     
\end{figure*}
 
\begin{figure*}
\begin{center} 
\stackon[5pt]{\includegraphics[width= 17.5cm, viewport=29 50 565 186, clip]{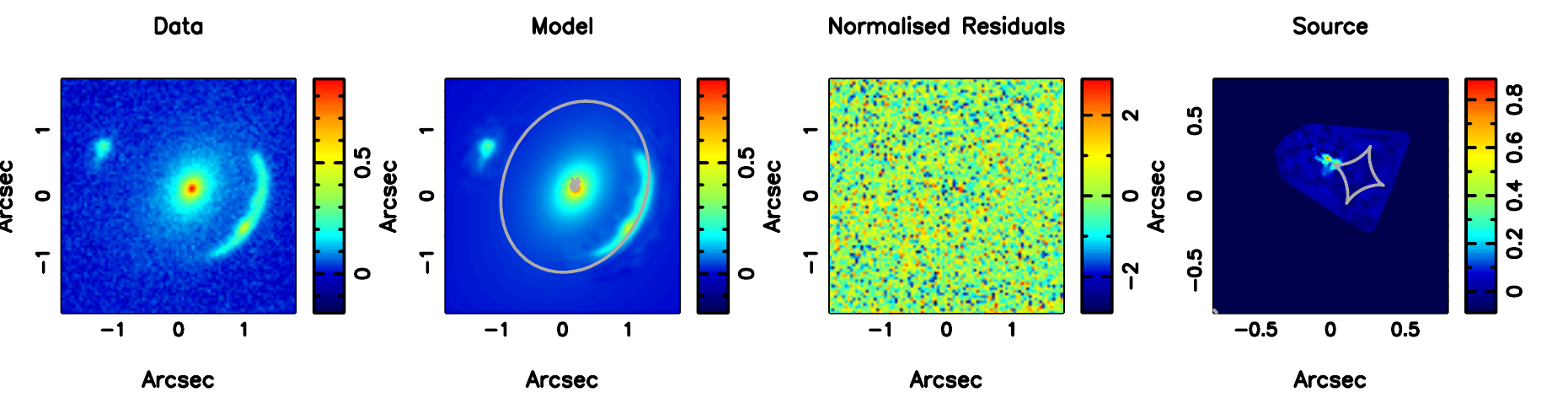}}{SDSS~J0742+3341}
\stackon[5pt]{\includegraphics[width= 17.5cm, viewport=29 50 565 186, clip]{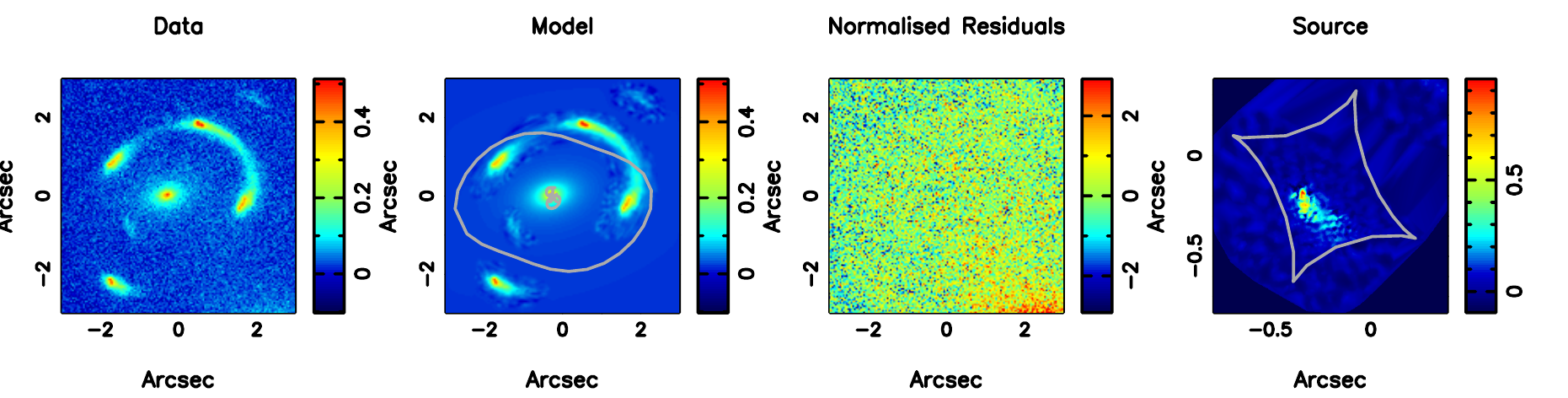}}{SDSS~J0755+3445}
\stackon[5pt]{\includegraphics[width= 17.5cm, viewport=29 50 565 186, clip]{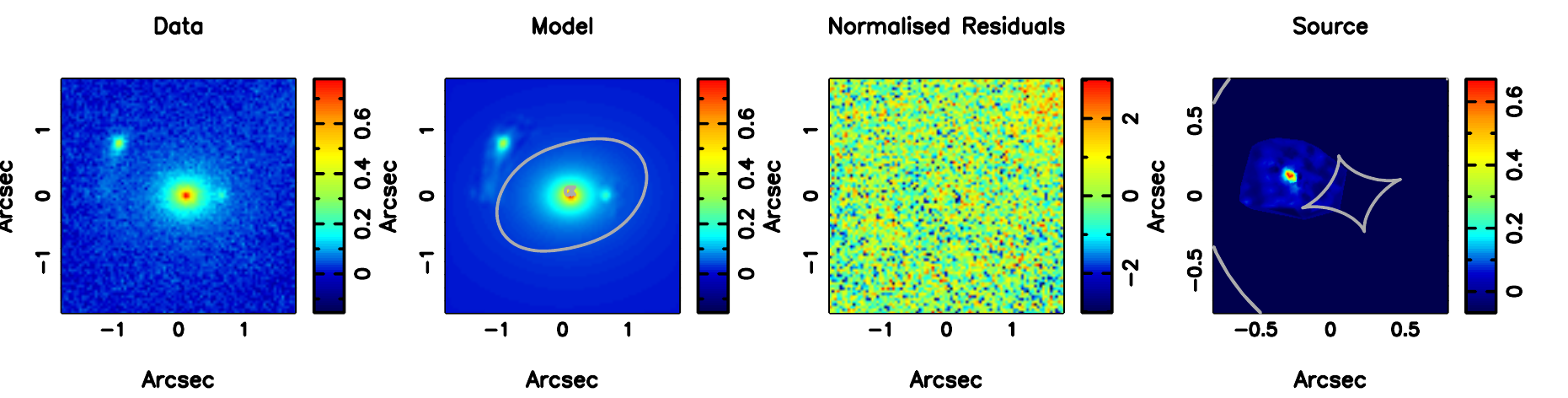}}{SDSS~J0856+2010}
\stackon[5pt]{\includegraphics[width= 17.5cm, viewport=29 50 565 186, clip]{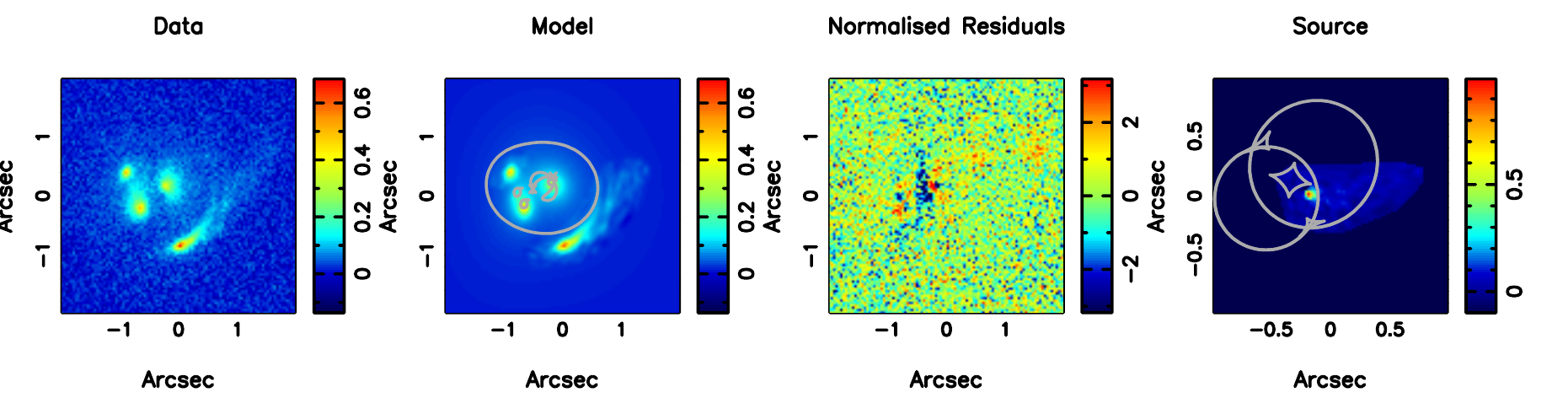}}{SDSS~J0918+4518}
\stackon[5pt]{\includegraphics[width= 17.5cm, viewport=29 50 565 186, clip]{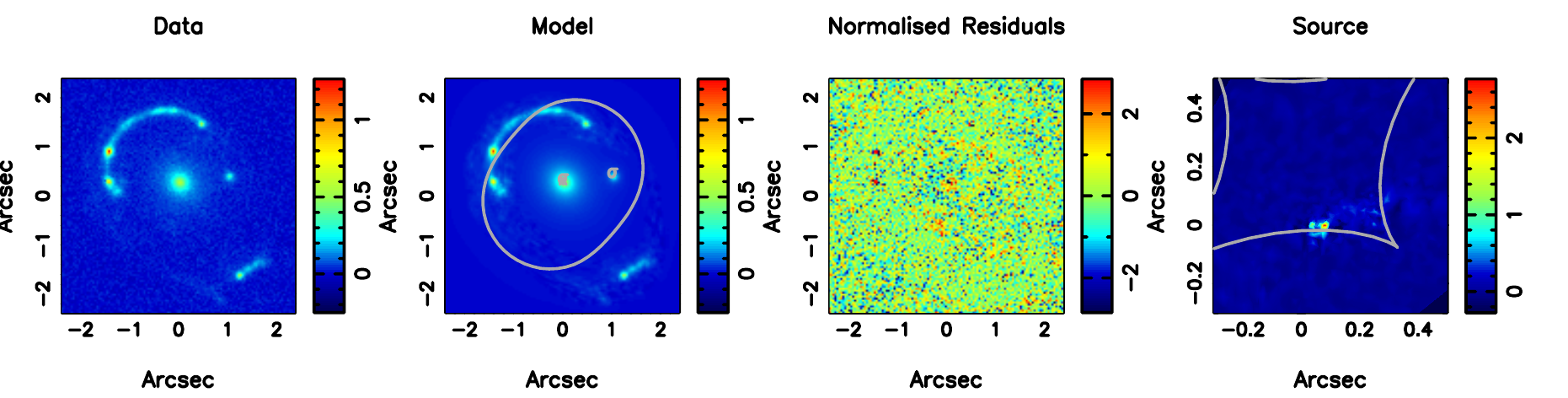}}{SDSS~J0918+5104}
\contcaption{}
\end{center}     
\end{figure*}

\begin{figure*}
\begin{center} 
\stackon[5pt]{\includegraphics[width= 17.5cm, viewport=29 50 565 186, clip]{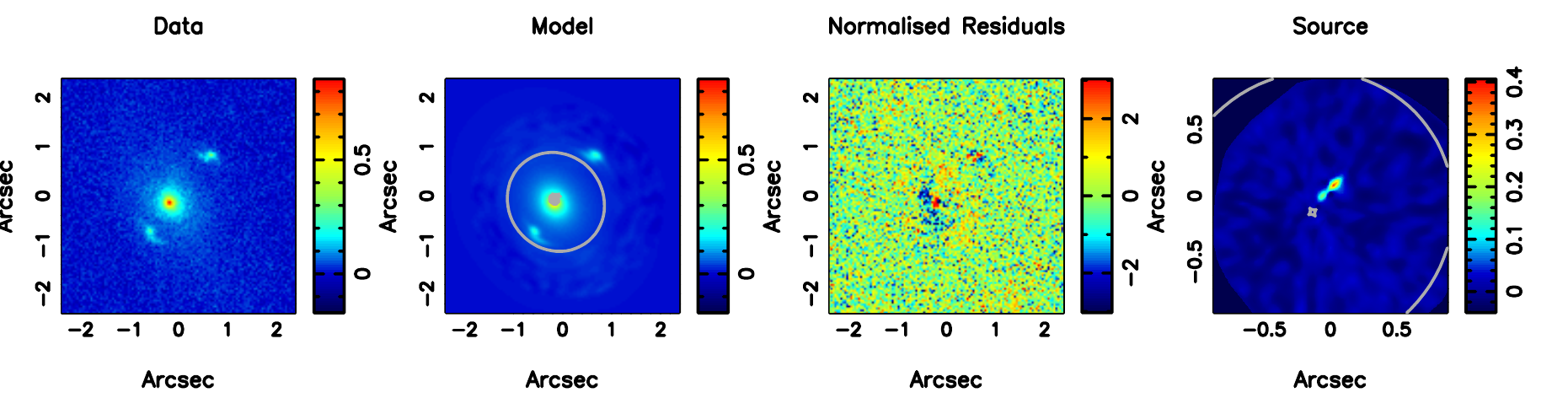}}{SDSS~J1110+2808}
\stackon[5pt]{\includegraphics[width= 17.5cm, viewport=29 50 565 186, clip]{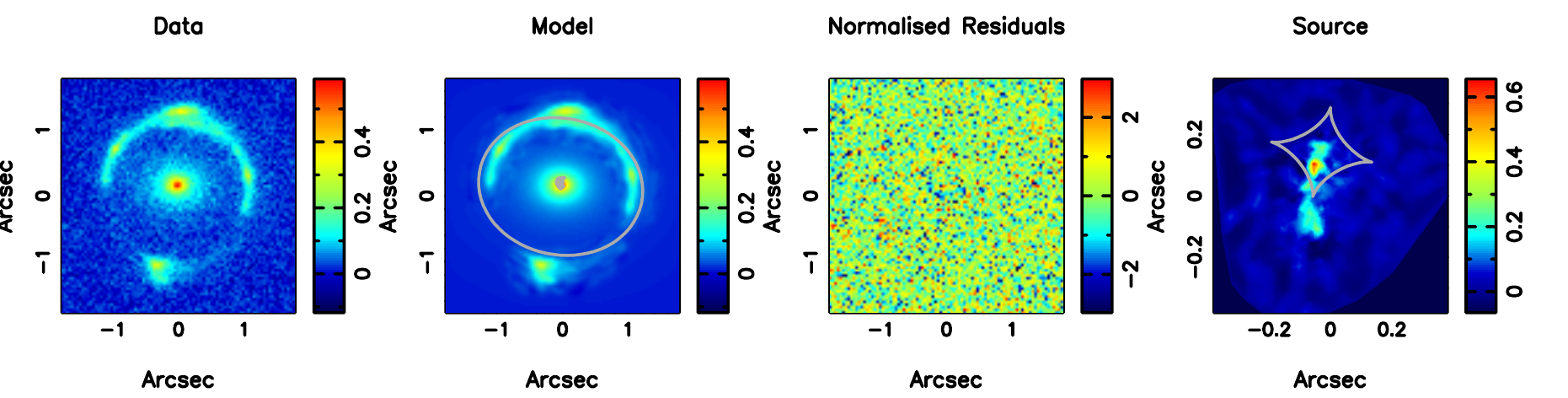}}{SDSS~J1110+3649}
\stackon[5pt]{\includegraphics[width= 17.5cm, viewport=29 50 565 186, clip]{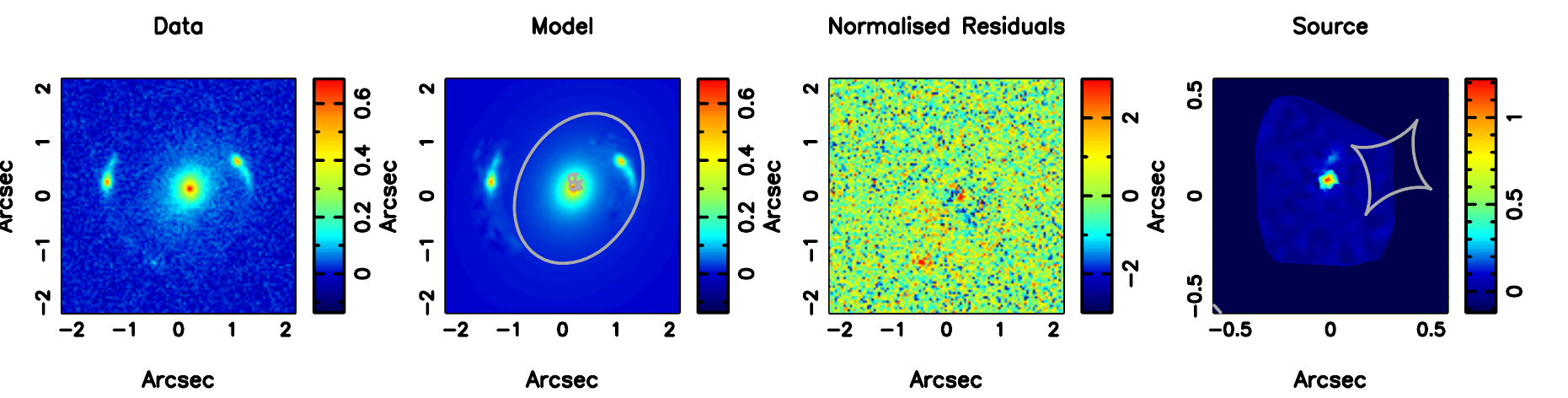}}{SDSS~J1141+2216}
\stackon[5pt]{\includegraphics[width= 17.5cm, viewport=29 50 565 186, clip]{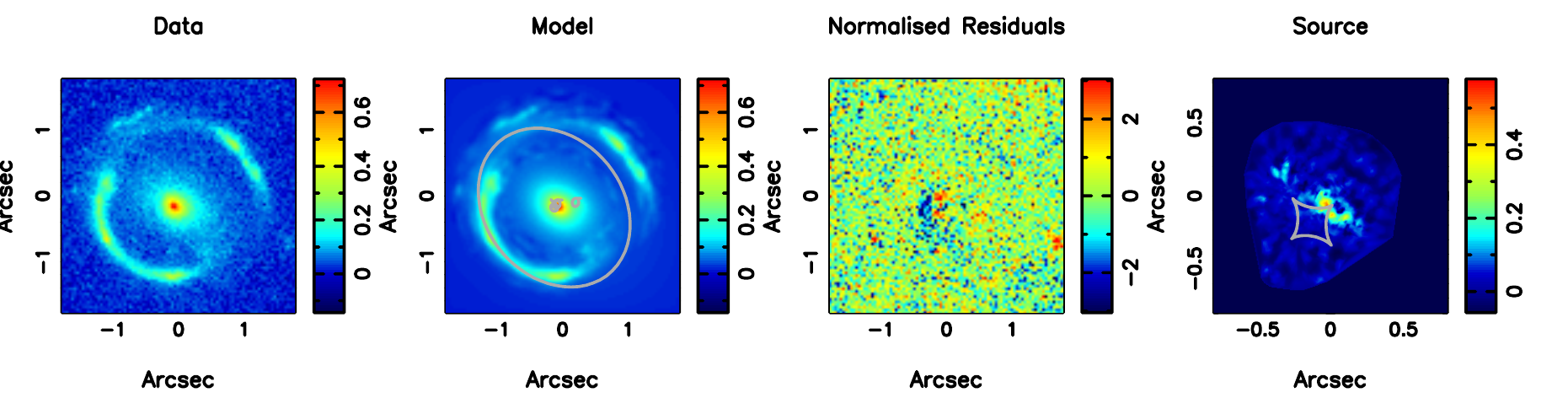}}{SDSS~J1201+4743}
\stackon[5pt]{\includegraphics[width= 17.5cm, viewport=29 50 565 186, clip]{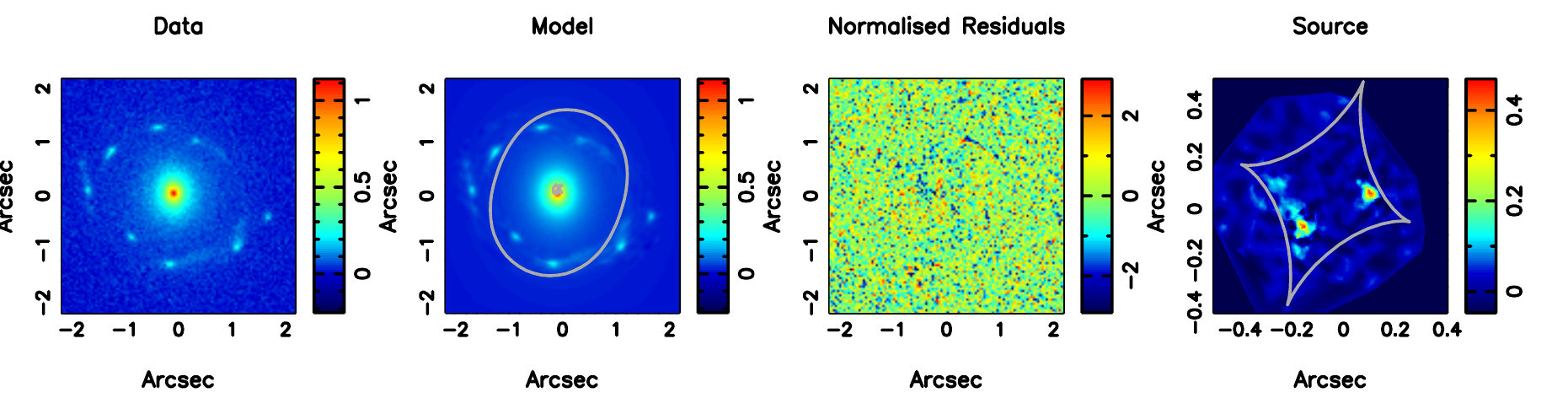}}{SDSS~J1226+5457}
\contcaption{}
\end{center}     
\end{figure*}
 
\begin{figure*}
\begin{center} 
\stackon[5pt]{\includegraphics[width= 17.5cm, viewport=29 50 565 186, clip]{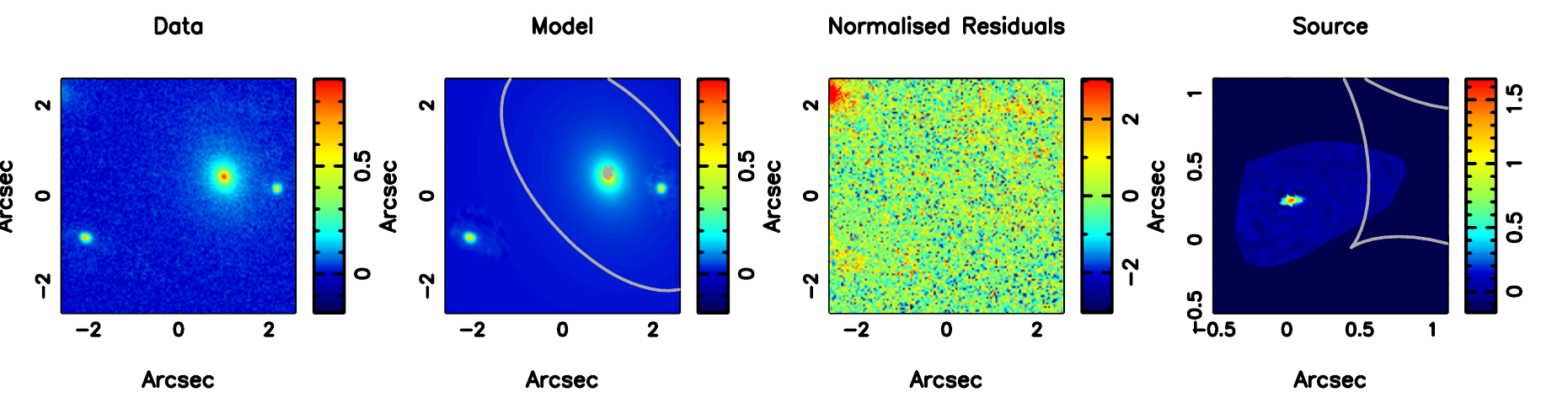}}{SDSS~J1529+4015}
\stackon[5pt]{\includegraphics[width= 17.5cm, viewport=29 50 565 186, clip]{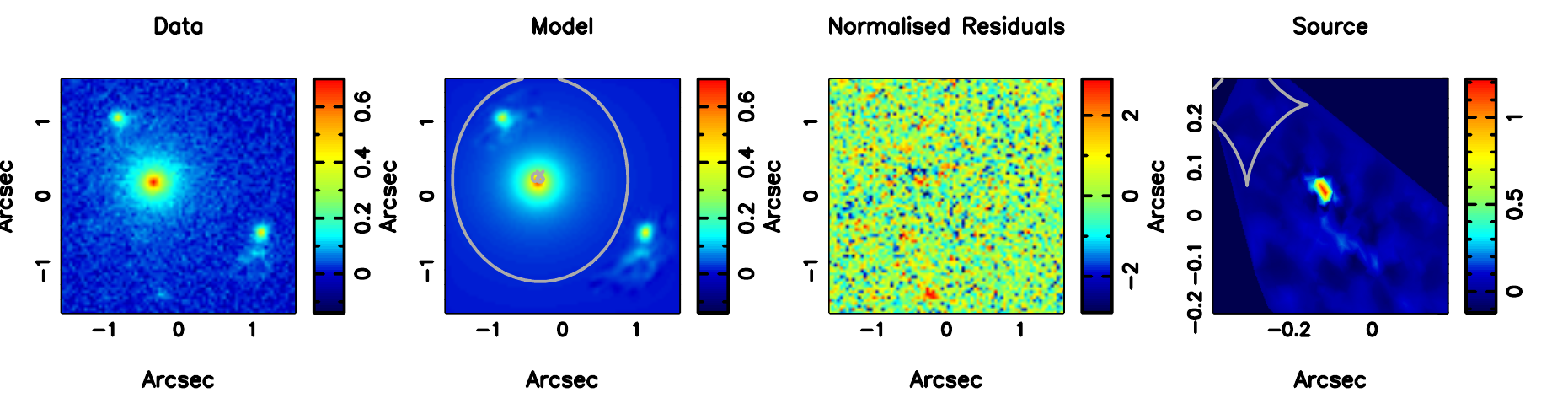}}{SDSS~J2228+2808}
\stackon[5pt]{\includegraphics[width= 17.5cm, viewport=29 50 565 186, clip]{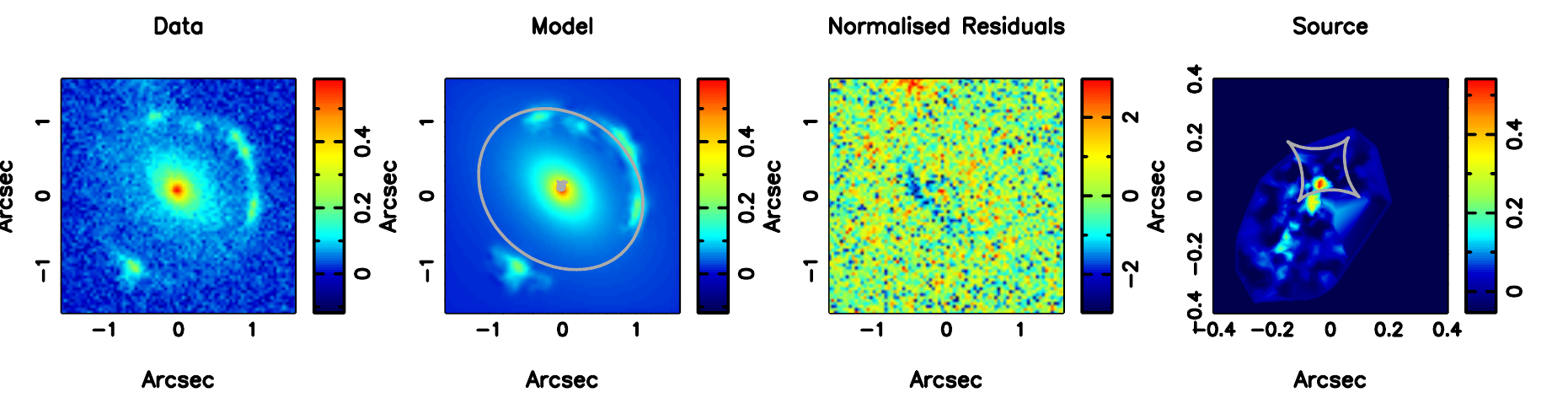}}{SDSS~J2342$-$0120}
\contcaption{}
\end{center}     
\end{figure*}

 \section{Intrinsic properties of LAE\texorpdfstring{\MakeLowercase{s}}{S}}
\label{sec:properties}

In this section, we present the inferred properties of the reconstructed LAE galaxies, including their morphologies, sizes, and SFR intensities. 

\subsection{Determining the intrinsic properties}

An essential application of gravitational lensing is to study the high-redshift Universe in a way that is not easily possible without the magnification provided by the foreground lensing galaxy. Potentially, this could be problematic, especially for extended objects, where the magnification can significantly change over the extent of the source. 

However, as our methodology already corrects for the distortion caused by the intervening gravitational lens (and the blurring caused by the PSF), our reconstructed sources are not affected by differential magnification. We note that due to the differing magnification over the source, there is an effective differential sensitivity and resolution across each object, although this is not an issue for our purposes. In fact we are interested in the global properties of these sources, rather than to their detailed structure. Therefore, we measure the intrinsic properties of the LAEs in the sample directly from their reconstructed UV surface brightness distributions.

To determine the star-forming properties of the sample, we use the standard methodology by assuming the rest-frame flux ($F_{\lambda}$) varies with some power-law, such that,
\begin{equation}
F_{\lambda} \propto \lambda^{\beta}
\end{equation}
where $\beta$ depends on the dust content of the galaxy and the age, metallicity and the initial mass function of the stellar population. As we have only one measurement of $F_{\lambda}$ from the F606W imaging, it is not possible to calculate $\beta$ for each object. Therefore, here we assume a typical value of $\beta = -1.5$ for LAEs at redshifts between 2 and 3 \citep{Nilsson09,Guaita10}. Note that as the central (rest-frame) wavelength of the data is close to the UV-reference wavelength of $\lambda_{\rm UV} = 1600$~{\AA} (see Table~\ref{tbl:list}), assuming a typical range of values of 0 to $-2.5$ for $\beta$ \citep[e.g.][]{Nilsson09,Blanc16,Hashimoto17} changes our derived values by at most 30 percent in the case of the integrated SFR of each object. If $\beta$ is not constant across the source, due to either a complex dust distribution or if there are multiple stellar populations (the former is the more likely for these sources; \citealt{Leitherer95,Bouwens09}), then the uncertainties in the SFR intensity maps from this assumption may be more pronounced. As we will compare with other samples similarly observed (at lower intrinsic angular resolutions), this will not significantly affect our analysis either. 

\begin{figure*}
\begin{center}
\includegraphics[width= 17.5cm]{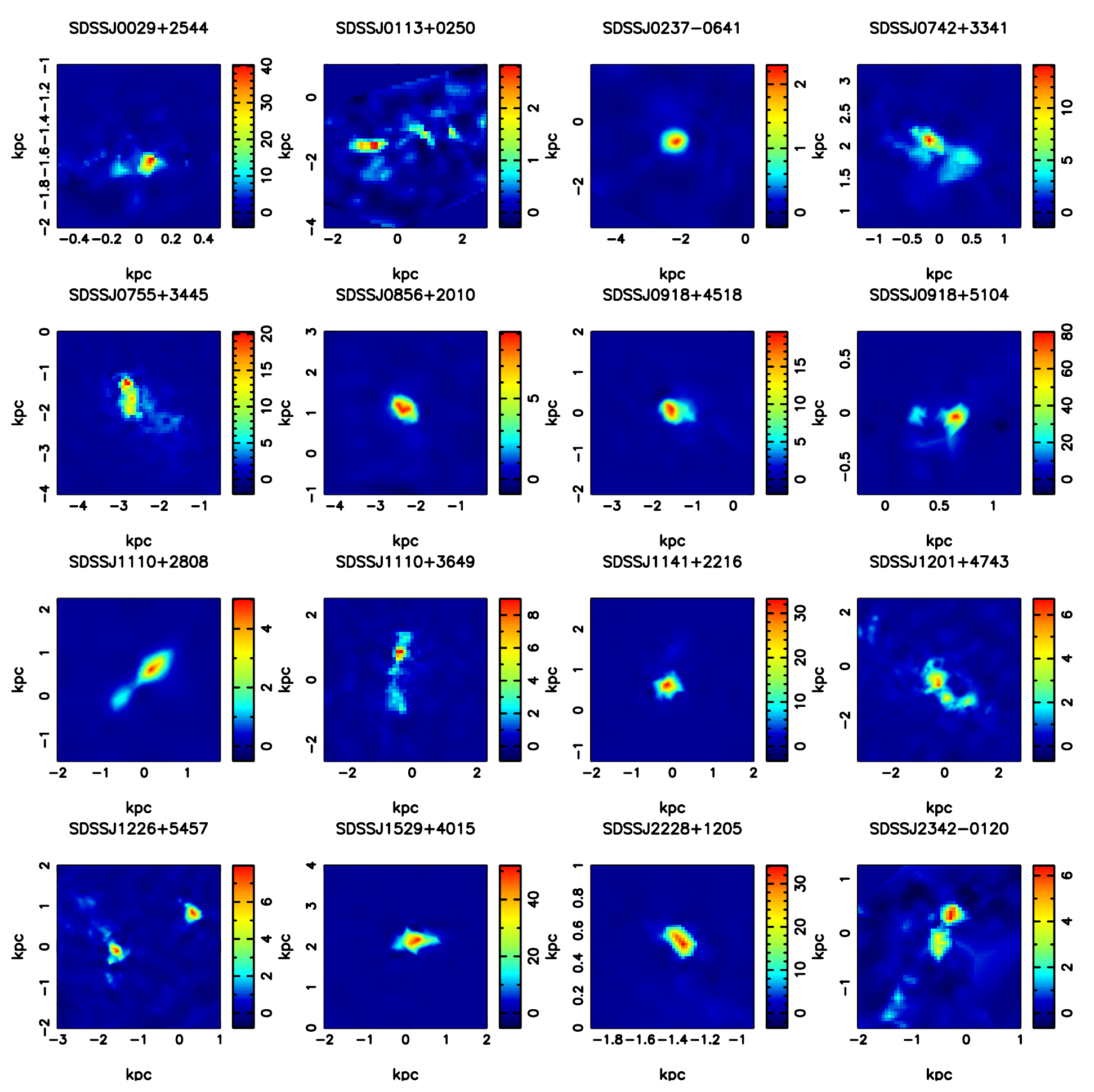}
\caption{The star-formation rate intensity of each object, based on the source reconstructions from the grid-based gravitational lens modelling. The colour-scale for each object is in units of M$_\odot$~yr$^{-1}$~kpc$^{-2}$.}
\label{fig:sfri}
\end{center}     
\end{figure*}

We compute the SFR of each LAE using,
\begin{equation}
{\rm SFR~[M_\odot~yr^{-1}]} = K_{\nu} \times L_{\nu},
\label{eq:SFR} 
\end{equation}
where $L_{\nu}$ is expressed in units of erg~s$^{-1}$~Hz$^{-1}$ and $K_{\nu}  = 1.15\times 10^{-28}$ is the FUV conversion factor (for the units given in equation \ref{eq:SFR}) between the spectral luminosity and ongoing SFR, adopted from \citet{Madau14}. Our maps of the SFR intensity (M$_\odot$~yr$^{-1}$~kpc$^{-2}$) of each object are shown in Fig.~\ref{fig:sfri}.

The reconstructed sources have a wide range of structures and morphologies, which we discuss in detail below. We note that \citet{Shu16c} have quantified the lens model parameters and the properties of their sources by fitting elliptical S\'ersic profiles to the reconstructed background galaxy surface brightness. They use pixellisation only for the multiple component sources in order to guess the number of S\'ersic components needed to describe the source surface brightness. However, we have found that in general the sources have quite an irregular surface brightness distribution and we, therefore, choose to characterise all the sources in the sample in a less model-dependent manner always using the pixellated reconstructions.

Our modelling procedure described in Section~\ref{sec:models} constrains the surface brightness at points on an \textit{irregular grid} defined by the mapping of the positions of the observed pixels back to the unlensed source plane \citep[see][for example]{V09}. We quantify the structure of each source by finding the minimum best-fitting ellipse that encloses the points within this grid that have a signal-to-noise ratio larger than 3, and we report the mean major-axis size and ellipticity of each ellipse in Table~\ref{tbl:SFR}. In addition we use the aperture defined by the ellipse to measure the integrated SFR of each object (note that as SDSS~J1226+5457 is composed of two well-separated objects, we report separate values for each of these components). The mean SFR intensities are determined by dividing these by the area of each elliptical region. We also report the \textit{peak} SFR intensity, which is robustly determined because gravitational lensing conserves the surface brightness and the multiple images allow for the effects of the PSF to be mitigated. Finally, we take one hundred random MCMC realisations of the lens mass model for each object and reconstruct the source in each case to quantify the errors on the morphological and physical properties of the source. We note that in some cases the zero uncertainties reported in Table~\ref{tbl:SFR} are due to insignificant changes from one MCMC realisation to the other. These properties are also presented in Table~\ref{tbl:SFR}, except for two cases, SDSS~J0201+3228 and SDSS~J0755+3445. The residuals in the image plane and the extremely clumpy structure of the source of SDSS~J0201+3228 suggest that multi-band data are needed to properly constrain the mass model and the source properties, as already discussed in Section~\ref{ssec:specific_cases}. In the case of SDSS~J0755+3445, we could not perform a MCMC to draw realisations and compute the errors for the reconstructed source since the lens model required pixellated potential corrections to obtain a good fit.

\subsection{Source morphologies}

\begin{figure}
\centering
\includegraphics[width=5.6cm,keepaspectratio,angle=-90]{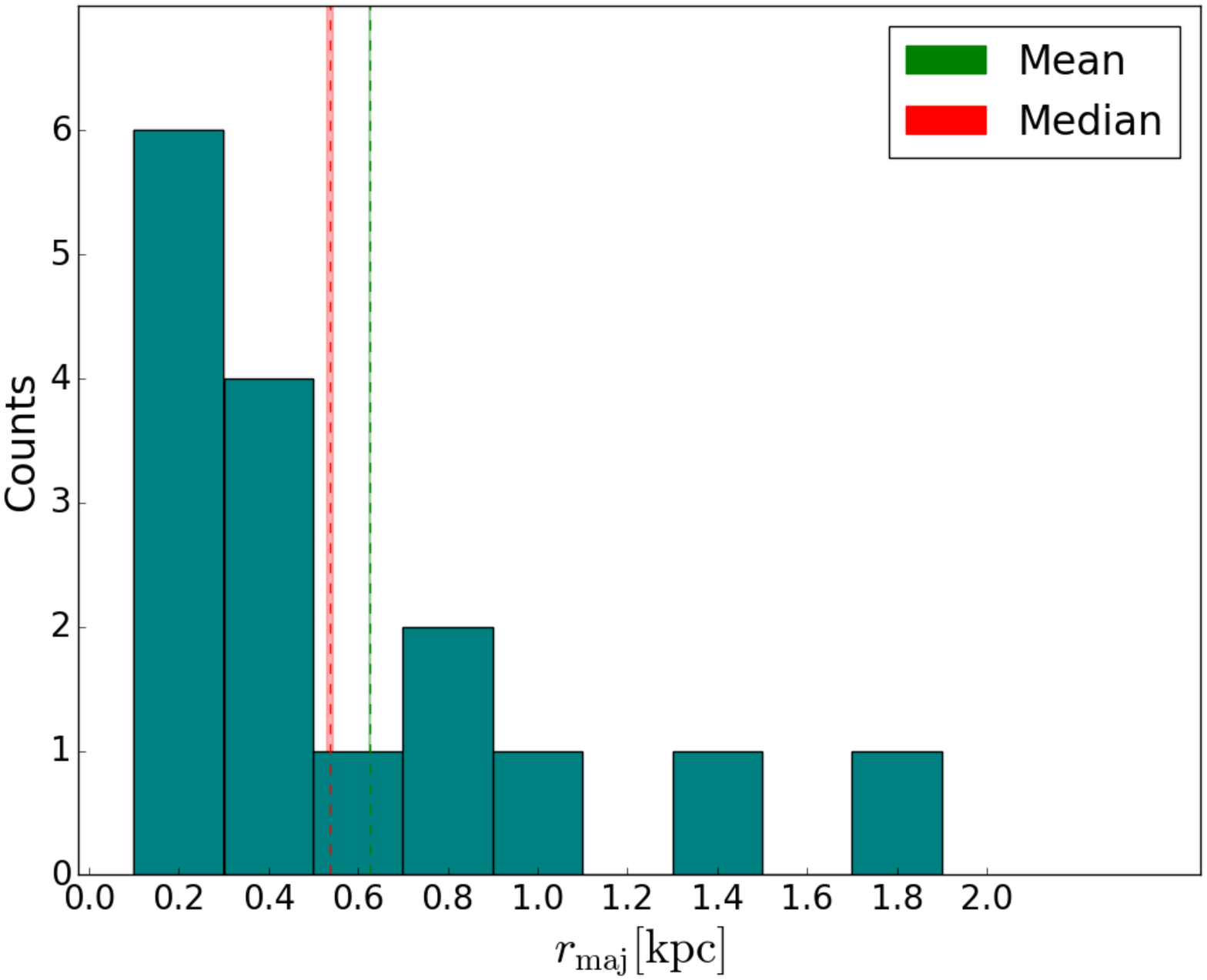}
\includegraphics[width=5.6cm,keepaspectratio,angle=-90]{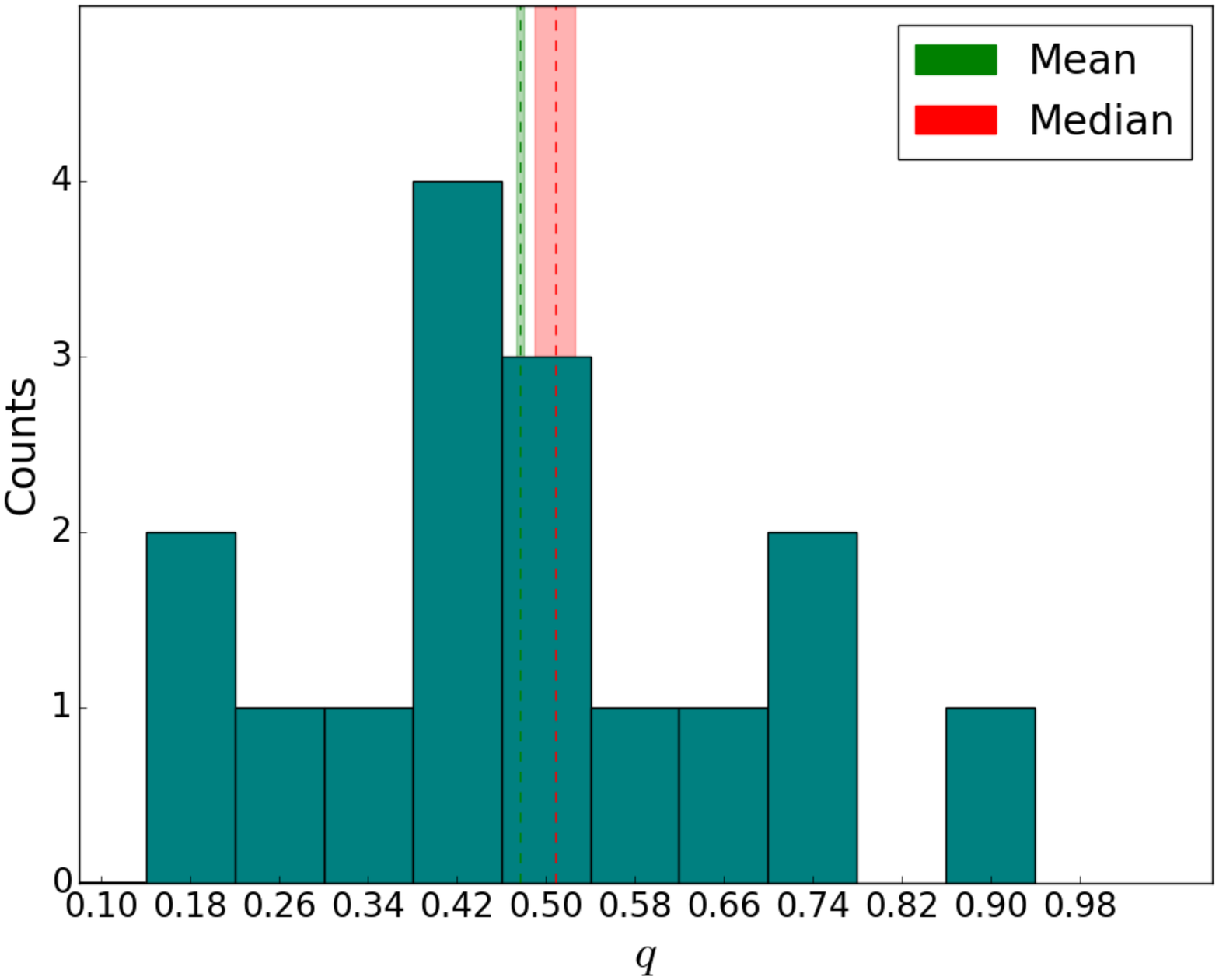}
\includegraphics[width=5.6cm,keepaspectratio,angle=-90]{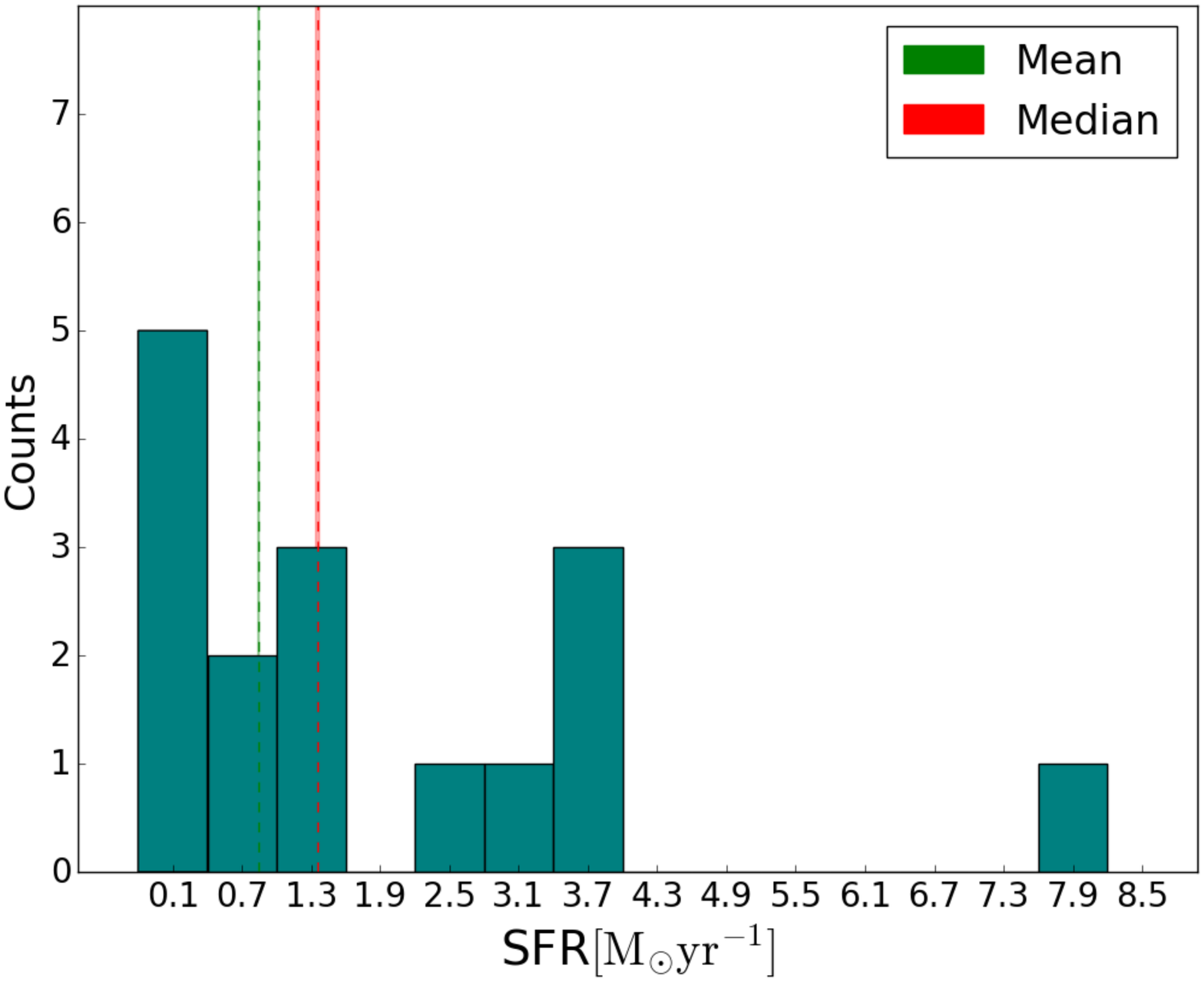}
\includegraphics[width=5.6cm,keepaspectratio,angle=-90]{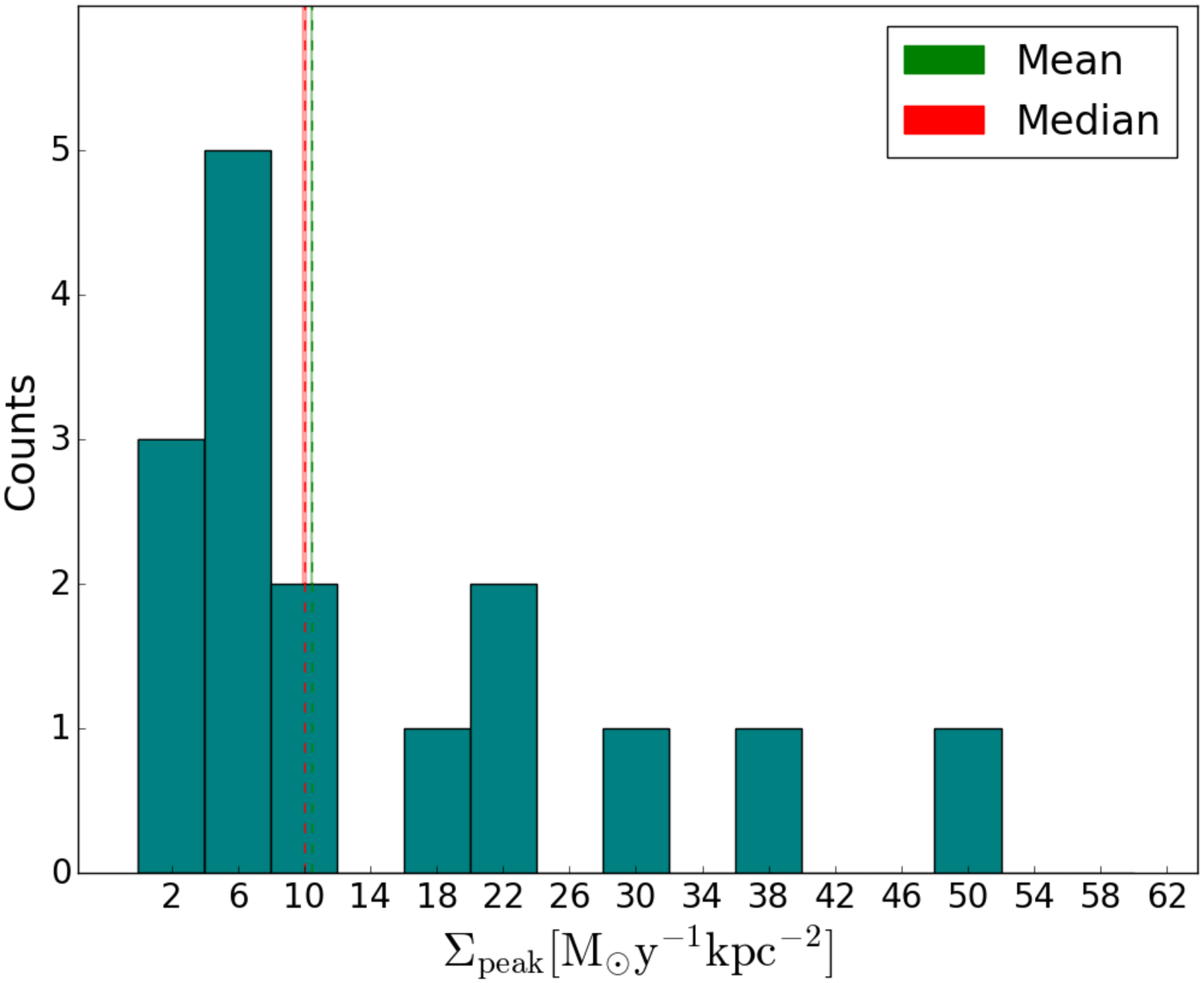}
\caption{From top to bottom: the distribution of major-axis sizes, axial-ratios, star-formation rates and peak star-formation rates intensities for the sample, with the mean and median (and their uncertainties) shown for reference.}
\label{fig:sfri-hist} 
\end{figure}

From Fig.~\ref{fig:sfri}, it is clear that the reconstructions of the LAEs have a range of morphological features, with multiple regions that have both compact and diffuse star-formation spread over 0.4 to 4 kpc in projected separation. Except for the double component source SDSS~J1226+5457, there tends to be a single dominant star-forming clump in each case. From Table~\ref{tbl:SFR} and Fig.~\ref{fig:sfri-hist}, we see that the major-axis sizes vary from 220~pc in the most compact sources (including low magnification doubly imaged systems where this is likely an upper limit; SDSS~J0237$-$0641 and SDSS~J0856+2010) to around 1.5 to 1.8 kpc in the cases of the most extended structures (SDSS~J0113+0250, SDSS~J1110+3649 and SDSS~J1201+4743). Note that in the most compact sources (major-axis $< 330$~pc), the intrinsic size is below the native pixel size of the UVIS camera on WFC3. This result highlights how gravitational lensing can be used to magnify structures in the high-redshift Universe; in fact, 10 out of 16 sources have major-axis sizes $<660$~pc (equivalent to 2 native pixel sizes), and the median major-axis size of the sample is 561$^{+13}_{-110}$~pc.

Also, almost all of the resolved sources have elongated structures that could be potentially interpreted as disks of star-formation. Again from Fig.~\ref{fig:sfri}, there is evidence that the multiple clumps of star-formation tend to be co-linear, as we would expect if there is a disk-like structure \citep{Swinbank09}. However, some kinematic tracer, such as emission lines in the optical- and mm-regime, will be needed to confirm this. We find that the axial-ratios of the ellipses that are fitted to the surface brightness distributions have a range from 0.18 in the most elongated case of SDSS~J0113+0250, to 0.91 for the almost circular case of SDSS J0918+4518. From Fig.~\ref{fig:sfri-hist}, we find that the distribution in axial ratio is quite flat, with a median of 0.51$^{+0.01}_{-0.01}$.

\subsection{Star-formation rates and intensities}

The derived SFRs from the UV-continuum for the LAEs in the sample vary by one order of magnitude, which gives some indication of the heterogeneous nature of the host galaxy properties. We recall that the objects in our sample were not selected based on their UV-continuum flux, but via the gravitational lensing of their Lyman-$\alpha$ emission, which may be affected by a varying Lyman-$\alpha$ escape fraction, slit-losses due to the finite size of the SDSS fibre and differential magnification. The lowest SFR in the sample is just 0.32~M$_{\odot}$~yr$^{-1}$ for SDSS~J0237$-$0641 and the largest SFR is 8.5~M$_{\odot}$~yr$^{-1}$ for SDSS~J1529+4015 (see Table \ref{tbl:SFR}). From Fig.~\ref{fig:sfri-hist}, we find that the distribution of SFRs is clustered towards the lower end, with 10 out of 16 sources having a SFR less than just 2~M$_{\odot}$~yr$^{-1}$. The median SFR for the whole sample is  1.37$^{+0.06}_{-0.07}$~M$_{\odot}$~yr$^{-1}$, which is comparable to the SFR of our own Milky Way.

Even though the galaxies are quite compact (median size of 561$^{+13}_{-110}$~pc; see above), their low SFRs still result in a relatively low average SFR intensity across each galaxy. We find that the average SFR intensities vary from 0.4 to 17.6~M$_{\odot}$~yr$^{-1}$~kpc$^{-2}$, with a median of 3.51$^{+0.07}_{-0.06}$~~M$_{\odot}$~yr$^{-1}$~kpc$^{-2}$. Furthermore, as described above, the morphology of the galaxies are typically characterised by a dominant star-forming component. The peak SFR intensity of the sample varies from 2.1 to 54.1~M$_{\odot}$~yr$^{-1}$~kpc$^{-2}$ (see Table \ref{tbl:SFR}). However, from Fig.~\ref{fig:sfri-hist} we find that the distribution has a peak around 6~M$_{\odot}$~yr$^{-1}$~kpc$^{-2}$ with an extended flat tail out to high peak SFR intensities. There does not seem to be a strong correlation between whether the LAE galaxy is compact or extended (based on the size of the major axis), and the peak in the SFR intensity. Overall, the sample has a median peak SFR intensity of 10.1$^{+0.5}_{-0.4}$~M$_{\odot}$~yr$^{-1}$~kpc$^{-2}$.

\begin{table*}
\caption{The derived AB F606W magnitude, major-axis size, flatness, star formation rate, mean star-formation-rate intensity, and peak star-formation-rate intensity for each LAE. Note that the SFR properties assume a continuum slope of $\beta = -1.5$. We expect the SFR to change by at most 30 percent for a typical range of $\beta$. The zero uncertainties are due to insignificant changes from one MCMC realisation to the other.}
\begin{tabular}{lcccrrr}
\hline
Name (SDSS) 		& m$_{\rm AB}$ & $r_{\rm maj}$ & $q$ & SFR & $<\Sigma>$ & $\Sigma_{\rm peak}$\\ 
& [mag] & [kpc] & & [M$_\odot$~yr$^{-1}$] & [M$_\odot$~yr$^{-1}$~kpc$^{-2}$] & [M$_\odot$~yr$^{-1}$~kpc$^{-2}$]\\
\hline
J0029$+$2544   &  27.05$^{+0.22}_{-0.13}$  &  0.410$^{+0.040}_{-0.044}$  &  0.57$^{+0.05}_{-0.06}$  &   0.96$^{+0.12}_{-0.18}$ & 3.33$^{+0.54}_{-0.65}$ & 30.85$^{+3.11}_{-2.95}$\\ 
J0113$+$0250   &  27.30$^{+0.08}_{-0.06}$  &  1.842$^{+0.230}_{-0.321}$  &  0.18$^{+0.02}_{-0.01}$  &   0.79$^{+0.04}_{-0.06}$ & 0.42$^{+0.13}_{-0.04}$ & 2.13$^{+0.28}_{-0.24}$\\
J0237$-$0641   &  28.18$^{+0.11}_{-0.18}$  &  0.224$^{+0.012}_{-0.011}$  &  0.81$^{+0.02}_{-0.02}$  &    0.32$^{+0.06}_{-0.02}$ & 2.50$^{+0.08}_{-0.03}$ & 4.22$^{+0.11}_{-0.09}$\\
J0742$+$3341   &  26.40$^{+0.06}_{-0.03}$  &  0.556$^{+0.011}_{-0.010}$  &  0.50$^{+0.01}_{-0.03}$  &   1.70$^{+0.05}_{-0.10}$ & 3.48$^{+0.02}_{-0.09}$ & 13.46$^{+0.23}_{-1.40}$\\
J0856$+$2010   &  27.30$^{+0.03}_{-0.01}$  &  0.242$^{+0.002}_{-0.012}$  &  0.58$^{+0.01}_{-0.01}$  &    0.71$^{+0.00}_{-0.02}$ & 6.69$^{+0.38}_{-0.05}$ & 11.54$^{+0.21}_{-0.05}$\\
J0918$+$4518   &  25.37$^{+0.04}_{-0.02}$  &  0.460$^{+0.013}_{-0.008}$  &  0.91$^{+0.02}_{-0.04}$  &   4.25$^{+0.11}_{-0.15}$ & 6.96$^{+0.16}_{-0.12}$ & 19.07$^{+0.43}_{-0.44}$\\
J0918$+$5104 &  25.94$^{+0.02}_{-0.02}$  &  0.372$^{+0.005}_{-0.004}$  &  0.66$^{+0.01}_{-0.01}$  &    2.63$^{+0.06}_{-0.06}$ & 9.21$^{+0.09}_{-0.13}$ & 54.10$^{+0.52}_{-0.30}$\\
J1110$+$2808   &  26.67$^{+0.02}_{-0.01}$  &  0.960$^{+0.010}_{-0.017}$  &  0.20$^{+0.01}_{-0.01}$  &   1.34$^{+0.02}_{-0.02}$ & 2.35$^{+0.03}_{-0.05}$ & 7.43$^{+0.17}_{-0.13}$\\ 
J1110$+$3649   &  25.57$^{+0.02}_{-0.05}$  &  1.520$^{+0.032}_{-0.016}$  &  0.31$^{+0.01}_{-0.00}$  &   3.79$^{+0.21}_{-0.08}$ & 1.69$^{+0.01}_{-0.01}$ & 8.94$^{+0.15}_{-0.15}$\\
J1141$+$2216   &  25.70$^{+0.05}_{-0.03}$  &  0.633$^{+0.019}_{-0.021}$  &  0.43$^{+0.01}_{-0.01}$  &   3.57$^{+0.15}_{-0.16}$ & 6.55$^{+0.05}_{-0.07}$ & 22.24$^{+0.11}_{-0.05}$\\
J1201$+$4743   &  25.36$^{+0.01}_{-0.02}$  &  1.067$^{+0.014}_{-0.012}$  &  0.55$^{+0.01}_{-0.02}$  &   4.09$^{+0.08}_{-0.04}$ & 2.09$^{+0.03}_{-0.03}$ & 6.81$^{+0.42}_{-0.10}$\\
J1226$+$5457A  &  27.66$^{+0.03}_{-0.06}$  &  0.575$^{+0.017}_{-0.010}$  &  0.49$^{+0.00}_{-0.06}$  &    0.58$^{+0.04}_{-0.01}$ & 1.15$^{+0.22}_{-0.01}$ & 7.22$^{+0.21}_{-0.08}$\\
J1226$+$5457B  &  27.83$^{+0.03}_{-0.06}$  &  0.242$^{+0.009}_{-0.005}$  &  0.76$^{+0.01}_{-0.01}$  &    0.50$^{+0.03}_{-0.01}$ & 3.55$^{+0.04}_{-0.04}$ & 6.79$^{+0.07}_{-0.06}$\\
J1529$+$4015   &  24.77$^{+0.02}_{-0.02}$  &  0.567$^{+0.011}_{-0.006}$  &  0.48$^{+0.01}_{-0.01}$  &  8.50$^{+0.17}_{-0.19}$ & 17.57$^{+0.19}_{-0.13}$ & 39.36$^{+0.59}_{-0.38}$\\
J2228$+$1205   &  26.66$^{+0.03}_{-0.03}$  &  0.279$^{+0.014}_{-0.006}$  &  0.49$^{+0.01}_{-0.02}$  &   1.50$^{+0.04}_{-0.04}$ & 12.41$^{+0.21}_{-0.28}$ & 24.69$^{+0.77}_{-0.24}$\\
J2342$-$0120   &  27.71$^{+0.05}_{-0.02}$  &  0.866$^{+0.002}_{-0.003}$  &  0.34$^{+0.00}_{-0.00}$  &    0.49$^{+0.01}_{-0.02}$ & 0.62$^{+0.01}_{-0.03}$ & 4.68$^{+0.04}_{-0.28}$\\
\hline
mean		& 26.13$^{+0.06}_{-0.05}$ &	0.649$^{+0.028}_{-0.032}$	& 0.48$^{+0.01}_{-0.02}$	&	0.86$^{+0.07}_{-0.08}$	&	2.04$^{+0.13}_{-0.11}$	& 10.54$^{+0.45}_{-0.44}$\\
median		& 26.66$^{+0.35}_{-0.03}$ &	0.561$^{+0.013}_{-0.110}$	& 0.51$^{+0.01}_{-0.01}$	&	1.37$^{+0.06}_{-0.07}$	&	3.51$^{+0.07}_{-0.05}$	& 10.1$^{+0.45}_{-0.44}$\\
\hline
\end{tabular}
\label{tbl:SFR} 
\end{table*}

\section{Discussion}
\label{sec:disc}

In this section, we discuss our results and compare the properties of the gravitationally lensed sample of LAEs investigated here with the large samples of non-lensed objects that have been studied thus far with different methods (e.g. narrowband imaging and spectroscopy). We also compare with the small number of lensed LAEs that have been studied.

\subsection{Morphology}
\label{ssection:morphology}

Current morphological studies of hundreds of non-lensed LAEs at moderate to high redshifts ($z\sim 2$ to 6) have been mainly limited by the angular resolution of the available imaging for both the line and continuum emitting regions, which could potentially bias the interpretation of data. For example, the UV-continuum morphologies have typically been classified as either single, marginally resolved components, or those with evidence for multiple components that are well separated on the sky. For those marginally resolved cases, studies have found that the de-convolved half-light radii of the single component LAEs are typically between 0.45 and 2 kpc in size, with a median half-light radius of 0.85 kpc \citep{Hagen16,Kobayashi16,Paulino-Afonso17}. In addition, there is strong evidence for a lack of evolution in the size of LAEs with redshift, suggesting that there is a characteristic scale for the stellar continuum emission within LAEs \citep{Malhotra12,Blanc16}. However, as the lower size limit approaches the pixel-scale of the imaging instruments on the {\it HST}, it is not clear how robust this conclusion is. Also, there are examples of LAEs with multiple individual UV-continuum components (with a similar size to those discussed above), but separated up to 4~kpc in projection \citep{Swinbank09,Kobayashi16}. In these cases, it is thought that the separate components are in the process of merging/interacting, which results in the triggering of star formation in the spatially distinct regions. However, these types of LAEs are somewhat rare, with an estimated merger fraction of just 10 to 30 percent, and so direct mergers are not thought to be the dominant mechanism for triggering star-formation in these galaxies \citep{Shibuya14}. From our reconstructed surface brightness distributions, we find that in more than half of the cases, the UV-continuum emission is made up of multiple compact components. Therefore, even if there is a well defined size for the H\,{\sc ii} regions within LAEs, there are multiple sites of star-formation that would appear to be a single marginally resolved component if these were non-lensed cases. Unfortunately, as we do not have resolved kinematic information, it is not possible to determine if these connected multiple components are part of a merging system or not (although, as we argue below, we believe that except for a few cases, they are part of disk systems, as opposed to kinematically independent components). However, we do find that one to two sources out of sixteen in our sample also show evidence of compact components that appear unconnected and are separated by up to 4 kpc, which is consistent with the observed merger rate of non-lensed LAEs. 

The gravitational lensing magnification provides higher intrinsic resolution for our reconstructed sources, and we find that the sample has a median size of $561_{-110}^{+13}$~pc, which is comparable to the lower limit on the size of the non-lensed samples of LAEs studied so far. However, we find that six sources in our sample have sizes below 0.5~kpc. The results from our analysis of the size scale of the UV-continuum are also consistent with the findings from studies of other lensed LAEs at high redshift, albeit with much smaller sample sizes. For example, \cite{Leclercq17}, expanding the sample analysed by \cite{Wisotzki16}, have found the minumim and maximum exponential scale-length of the UV continuum to be 0.11 kpc and 1.58 kpc respectively, in agreement with what previously found by \cite{Wisotzki16} for the smaller sample. \cite{Patricio16} carried out an analysis of a lensed star-forming galaxy at $z=3.5$ and estimated a size of 0.34 kpc for the UV-continuum. Unfortunately, we cannot directly compare our results with theirs as we have used a different definition of size. This suggests that previous studies of non-lensed LAEs may have been biased by the angular resolution of the data and therefore it is not clear whether there really is limited size evolution with redshift for the UV-continuum emission of LAEs.

The size and structure of the UV-continuum is also thought to be related to the observed Lyman-$\alpha$ emission in these galaxies. Overall, there is evidence that the Lyman-$\alpha$ is more extended than the UV-continuum by factors of a few, and there is evidence for Lyman-$\alpha$ haloes that are both offset from the UV-continuum and extended by a few to 10~kpc \citep{Momose14}. For example, \cite{Leclercq17} found that Lyman-$\alpha$ haloes have sizes from 4 to 20 times bigger than the UV continuum scale-length, with a median size ratio of 10.8, in agreement with \cite{Wisotzki16}. The implications for this are that either the surface brightness of any extended UV-continuum is too low to be detected, relative to the Ly$\alpha$, or that the compact star-forming regions traced by the UV-continuum are ionising the surrounding gas up to several kpc away. Detecting any low surface brightness continuum emission would also determine whether there is diffuse ongoing star-formation over a larger extent than thought, as opposed to there being a few compact regions of intense star-formation that may or not be related to the Lyman-$\alpha$ sources; the latter would again suggest that the star-forming galaxy and the LAE are independent systems that are interacting. Again, our reconstructed surface brightness distributions show compact and diffuse (and lower surface brightness) UV-emission, which would favour the interpretation that we are observing extended star-forming galaxies with multiple compact H\,{\sc ii} regions that are ionising the surrounding gas and therefore Lyman Continuum Leakers. This result is in agreement with what found by \cite{Wisotzki16}, \cite{Hashimoto17} and \cite{Leclercq17}. Resolved imaging of the Lyman-$\alpha$ emission will be needed to confirm this conclusion.

Also, it has been suggested that the Lyman-$\alpha$ equivalent width (EW) is connected to the morphology and size of the UV-continuum, with more optically compact LAEs having a much larger EW \citep{Law12}. This has been interpreted as evidence for the galaxy morphology having an impact on whether Lyman-$\alpha$ is produced and whether it is able to escape from the host galaxy inter-galactic medium. However, it has also been reported that there is no correlation between the half-light radius and the EW \citep{Wisotzki16,Hashimoto17,Leclercq17}, which may be due to the LAEs having a diverse set of host galaxy properties \citep{Bond12}, or from the observational biases associated with narrowband observations of the EW from such compact sources \citep{Oyarzun17}. There is also evidence for an anti-correlation between the ellipticity of the UV-continuum and the EW \citep{Shibuya14}, and a correlation between the ellipticity and size of the UV-continuum and the detection of Lyman-$\alpha$ \citep{Kobayashi16,Hagen16}. 

The implication of such observations is that the Lyman-$\alpha$ escape fraction is not a function of the ellipticity, and hence, the inclination angle of the star-forming disk, contrary to radiative transfer models in hydrodynamical galaxy formation simulations that predict an anisotropic escape path for the Lyman-$\alpha$ \citep{Verhamme12}. Such models require the line-of-sight through the star-forming disk to be optically thick, with the Lyman-$\alpha$ emission being preferentially seen perpendicular to the star-forming disk (i.e. face-on objects that are characterised with a low ellipticity). Instead, the current observational data for non-lensed samples of LAEs favour a clumpy model for the star-forming regions, where the sight-lines for Lyman-$\alpha$ to escape are distributed randomly within a galaxy \citep{Gronke14}. Therefore the importance of ellipticity for the Lyman-$\alpha$ escape fraction is still unclear and debated, with the data suggesting the absence of any correlation among the two. 

\begin{figure}
\centering
\includegraphics[width=7.6cm,viewport=30 180 545 600, clip]{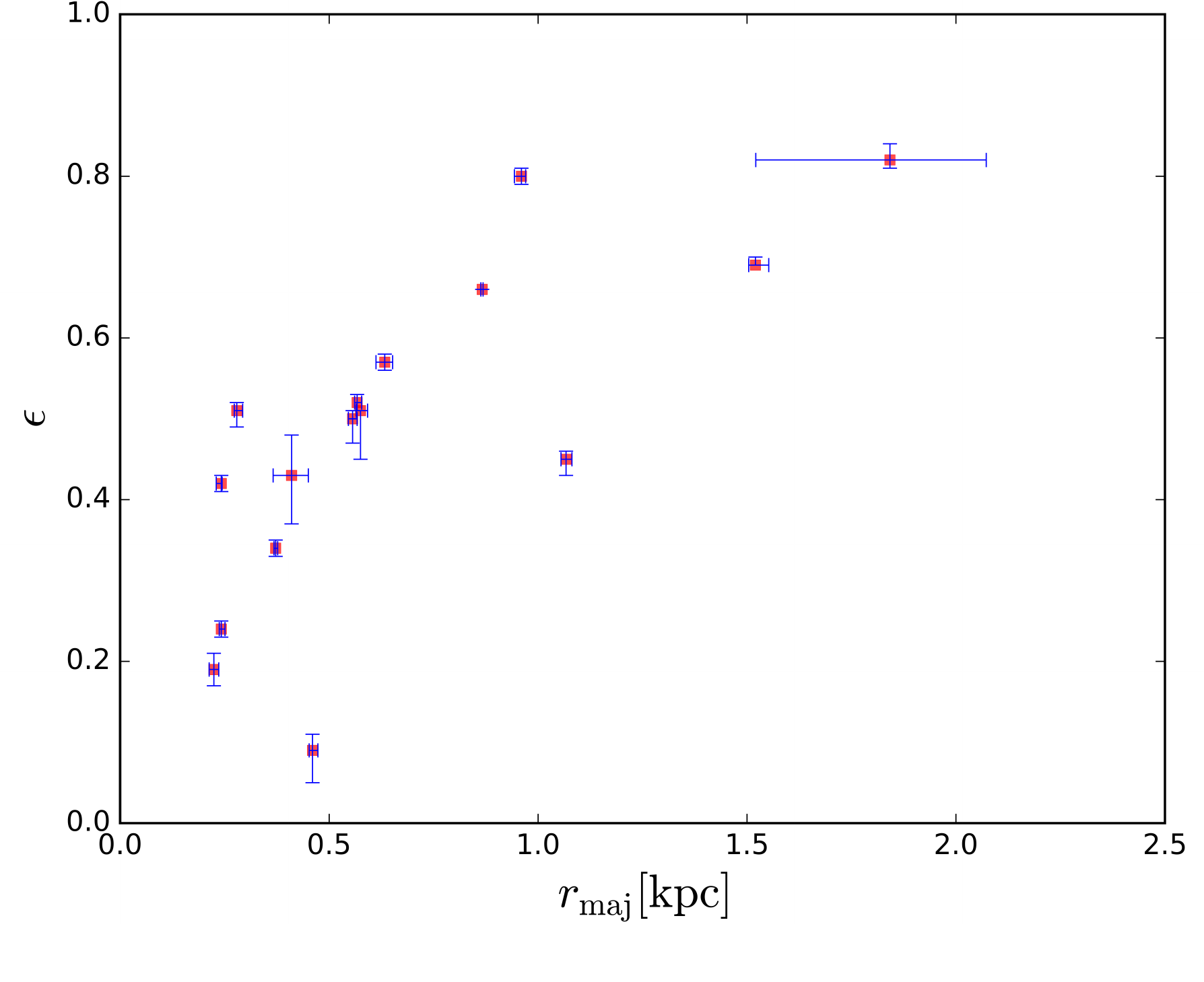}
\caption{The size of the reconstrcuted sources in the sample against their ellipticity ($\epsilon = 1-q$, where $q$ is reported in Table~\ref{tbl:SFR}).}
\label{fig:r_vs_e} 
\end{figure}

The morphology of our reconstructed LAEs support the results found at lower angular resolution for the non-lensed samples. There is a clear positive correlation between the size and the ellipticity of the UV-continuum emission (see Fig.~\ref{fig:r_vs_e}), at least for $\epsilon > 0.2$. In addition, the sources are also found to be highly elliptical, with more than half of the sample having axis-ratios less than 0.6, which is consistent with disk-like morphologies. Therefore, our data are consistent with the expectations from the clumpy model for the Lyman-$\alpha$ escape sight-lines. Unfortunately, there is no available imaging for the full Lyman-$\alpha$ emission for this sample. Therefore, we cannot currently measure the total Lyman$-\alpha$ fluxes and EWs, without making assumptions on the spectral slope and the total extension of the Lyman-$\alpha$ relative to the continuum.

\subsection{Star formation rate}
\label{ssection:SFR}

Many efforts have been made to estimate the SFR intensities of LAEs at different redshifts, as this is needed to understand their role in the build up of galaxies over cosmic time. Several samples consisting of hundreds of LAEs at $z\sim$2 to 4 have been analysed via stacking methods, finding a range of SFR of 1.5 up to 11~M$_\odot$~yr$^{-1}$ based on the luminosity of the UV-continuum of various samples \citep{Blanc16,Nakajima12,Alavi14,Kusakabe15}. The largest uncertainty in these analyses is determining the dust extinction, which requires multiple broad band filters to measure the reddening and is also sensitive to the assumptions about the extinction law for these galaxies, although the effect of dust is expected to decrease substantially for $z>2$ \citep{Smith18}. However, it is expected that the reddening is between $E(B-V) = 0$ to 0.5 \citep{Blanc16,Matthee16}, with a typical attenuation of about 3 for $\beta = -1.5$ and assuming a \citet{Calzetti00} extinction law. For consistency, we compare our results to the non-dust corrected SFRs of these samples of non-lensed LAEs, but they are likely around 3 times higher than stated here when the dust attenuation is taken into account.

Overall, we find that the median SFR of the galaxies in the lensed sample of LAEs studied here is lower, but certainly comparable, to those obtained from the stacking techniques, with a median SFR of 1.4~M$_\odot$~yr$^{-1}$. Such stacking methods can be biased by outliers, however, we find that from the direct measurement of the SFR of individual lensed galaxies that the median SFR is similar and therefore, the stacking methods are likely tracing the general properties of LAEs.

\section{Summary}
\label{sec:conc}

We have analysed the first statistically significant sample of LAEs at $z\sim2.5$ that are gravitationally lensed by massive foreground galaxies at $z\sim0.5$. By studying LAEs that are intrinsically extended, the gravitational lensing signature typically consists of extended arcs and Einstein rings, which give a large number of constraints to the lensing mass model and allow the effect of the PSF to be de-convolved and the intrinsic source structures to be obtained.

We have modelled the gravitational potential of all of the systems in the sample using a power-law mass distribution and an external shear, with additional small-scale corrections to the potential where required, and simultaneously inferred the surface brightness distribution for the lensing galaxies. This modelling is more sophisticated than has been carried out so far and provides the most robust inferences of the reconstructed sources. We have found that most of the deflectors are well described by a power-law mass-density profile, which is close to being isothermal with $<\gamma '>=2.00\pm0.01$. Except for a few cases where significant residuals are present in correspondence of the lensing galaxy, the light distribution is well represented by one or two S\'ersic components. All of our models, except for a few peaks in the brightness of the lensed images, have residuals which are at the noise level. Most of our lens-mass models agree with those reported by S16, but often present a lower level of image residuals, which in part is likely due to the way the source is solved for with our technique, but is also due to subtle differences in the lens macro-model.

We have studied the intrinsic properties of the UV-continuum of the reconstructed LAEs at $z\sim2.5$, finding that they have a median integrated SFR of $1.4^{+0.06}_{-0.07}$~M$_\odot$~yr$^{-1}$ and a peak SFR intensity between 2.1 to 54.1~M$_{\odot}$~yr$^{-1}$~kpc$^{-2}$. These galaxies are quite compact with a range of radii from 0.2 to 1.8~kpc and a median size of 561$^{+13}_{-110}$~pc (semi-major axis), and complex morphologies with several compact and diffuse components separated by 0.4 to 4 kpc \citep[see also][]{Cornachione18}. Our lower limit to the intrinsic size is about a factor of two smaller than that found for non-lensed LAEs, which highlights the power of gravitational lensing and sophisticated lens modelling techniques for resolving such objects in the high redshift Universe. Most importantly, we find that the LAEs are quite elliptical, with a median axial-ratio of $0.51\pm0.01$. This morphology is consistent with disk-like structures of star-formation for 76 percent of the sample, which would rule out models where the Lyman-$\alpha$ emission is only seen perpendicular to the disk, and favours those clumpy models for the escape lines-of-sight for Ly$\alpha$. This is in agreement with the results for non-lensed LAEs studied at similar redshifts, but is more robust given the improved angular resolution of the our analysis. We also find that 14 percent of the sample have off-axis components that may be associated with mergers, which is again consistent with the analysis of non-lensed LAEs. Overall, the morphologies of the LAEs in our sample agree quite well with the results in the literature, where the effective angular resolution is less and stacking techniques are needed.

Resolved kinematical information will be needed to confirm the disk-like nature and possible merger scenario for the LAEs in the sample, and to study the distribution of the ionised gas with respect to the UV-continuum sources. Such observations will also allow a robust resolved study of the Lyman-$\alpha$ escape fraction for these objects, as the current low resolution data from SDSS are limited by slit-losses. Also, as the UV-continuum and the Lyman-$\alpha$ are expected to be offset, it is not possible to use the magnifications derived here to probe the intrinsic luminosity of the Lyman-$\alpha$ emission. Generally, our analysis further demonstrates gravitational lensing as a powerful tool to analyse and resolve the detailed structure of high-redshift galaxies, allowing the study of their physical and morphological properties at a resolution otherwise only achievable with nearby targets.  

\section*{Acknowledgments}
The work presented in this paper is based on observations made with the NASA/ESA Hubble Space Telescope, and obtained from the Data Archive at the Space Telescope Science Institute, which is operated by the Association of Universities for Research in Astronomy, Inc., under NASA contract NAS 5-26555. These observations are associated with program 14189. SV has received funding from the European Research Council (ERC) under the European Union's Horizon 2020 research and innovation programme (grant agreement No 758853).

\bibliographystyle{mnras}
\bibliography{ms}

\bsp	
\label{lastpage}
\end{document}